\documentclass{aa}
\makeatletter
\def\@floatboxreset{\@floatpenalty\@M\centering}
\makeatother
\usepackage[stable]{footmisc}
\usepackage{amsmath}
\usepackage{amsfonts}
\usepackage{textgreek}

\usepackage[varg]{txfonts}
\usepackage{placeins}
\usepackage{natbib}
\bibpunct{(}{)}{;}{a}{}{,}
\usepackage{graphicx}
\usepackage{epstopdf}
\usepackage{ifthen}
\usepackage{mathtools}
\usepackage[breaklinks, colorlinks, citecolor=blue]{hyperref}
\usepackage{url}
\usepackage{hyperref}
\usepackage[table,dvipsnames]{xcolor}
\usepackage{physics}

\usepackage{silence}
\def\setsymbol#1#2{\expandafter\def\csname #1\endcsname{#2}}
\def\getsymbol#1{\csname #1\endcsname}

\def\Planck{\textit{Planck}}

\newbox\tablebox    \newdimen\tablewidth
\def\leaderfil{\leaders\hbox to 5pt{\hss.\hss}\hfil}
\def\endPlancktable{\tablewidth=\columnwidth 
    $$\hss\copy\tablebox\hss$$
    \vskip-\lastskip\vskip -2pt}

\def\tablenote#1 #2\par{\begingroup \parindent=0.8em
    \abovedisplayshortskip=0pt\belowdisplayshortskip=0pt
    \noindent
    $$\hss\vbox{\hsize\tablewidth \hangindent=\parindent \hangafter=1 \noindent
    \hbox to \parindent{$^#1$\hss}\strut#2\strut\par}\hss$$
    \endgroup}
\def\doubleline{\vskip 3pt\hrule \vskip 1.5pt \hrule \vskip 5pt}

\def\L2{\ifmmode L_2\else $L_2$\fi}

\def\DeltaT{\ifmmode \Delta T\else $\Delta T$\fi}
\def\deltat{\ifmmode \Delta t\else $\Delta t$\fi}
\def\fknee{\ifmmode f_{\rm knee}\else $f_{\rm knee}$\fi}
\def\Fmax{\ifmmode F_{\rm max}\else $F_{\rm max}$\fi}
\def\solar{\ifmmode{\rm M}_{\mathord\odot}\else${\rm M}_{\mathord\odot}$\fi}
\def\Msolar{\ifmmode{\rm M}_{\mathord\odot}\else${\rm M}_{\mathord\odot}$\fi}
\def\Lsolar{\ifmmode{\rm L}_{\mathord\odot}\else${\rm L}_{\mathord\odot}$\fi}
\def\inv{\ifmmode^{-1}\else$^{-1}$\fi}
\def\mo{\ifmmode^{-1}\else$^{-1}$\fi}
\def\sup#1{\ifmmode ^{\rm #1}\else $^{\rm #1}$\fi}
\def\expo#1{\ifmmode \times 10^{#1}\else $\times 10^{#1}$\fi}
\def\,{\thinspace}
\def\lsim{\mathrel{\raise .4ex\hbox{\rlap{$<$}\lower 1.2ex\hbox{$\sim$}}}}
\def\gsim{\mathrel{\raise .4ex\hbox{\rlap{$>$}\lower 1.2ex\hbox{$\sim$}}}}

\def\simprop{\mathrel{\raise .4ex\hbox{\rlap{$\propto$}\lower 1.2ex\hbox{$\sim$}}}}
\def\deg{\ifmmode^\circ\else$^\circ$\fi}
\def\pdeg{\ifmmode $\setbox0=\hbox{$^{\circ}$}\rlap{\hskip.11\wd0 .}$^{\circ}
          \else \setbox0=\hbox{$^{\circ}$}\rlap{\hskip.11\wd0 .}$^{\circ}$\fi}
\def\arcs{\ifmmode {^{\scriptstyle\prime\prime}}
          \else $^{\scriptstyle\prime\prime}$\fi}
\def\arcm{\ifmmode {^{\scriptstyle\prime}}
          \else $^{\scriptstyle\prime}$\fi}
\newdimen\sa  \newdimen\sb
\def\parcs{\sa=.07em \sb=.03em
     \ifmmode \hbox{\rlap{.}}^{\scriptstyle\prime\kern -\sb\prime}\hbox{\kern -\sa}
     \else \rlap{.}$^{\scriptstyle\prime\kern -\sb\prime}$\kern -\sa\fi}
\def\parcm{\sa=.08em \sb=.03em
     \ifmmode \hbox{\rlap{.}\kern\sa}^{\scriptstyle\prime}\hbox{\kern-\sb}
     \else \rlap{.}\kern\sa$^{\scriptstyle\prime}$\kern-\sb\fi}
\def\ra[#1 #2 #3.#4]{#1\sup{h}#2\sup{m}#3\sup{s}\llap.#4}
\def\dec[#1 #2 #3.#4]{#1\deg#2\arcm#3\arcs\llap.#4}
\def\deco[#1 #2 #3]{#1\deg#2\arcm#3\arcs}
\def\rra[#1 #2]{#1\sup{h}#2\sup{m}}

\def\dots{\relax\ifmmode \ldots\else $\ldots$\fi}
\def\WHzsr{\ifmmode $W\,Hz\mo\,sr\mo$\else W\,Hz\mo\,sr\mo\fi}
\def\mHz{\ifmmode $\,mHz$\else \,mHz\fi}
\def\GHz{\ifmmode $\,GHz$\else \,GHz\fi}
\def\mKs{\ifmmode $\,mK\,s$^{1/2}\else \,mK\,s$^{1/2}$\fi}
\def\muKs{\ifmmode \,\mu$K\,s$^{1/2}\else \,$\mu$K\,s$^{1/2}$\fi}
\def\muKRJs{\ifmmode \,\mu$K$_{\rm RJ}$\,s$^{1/2}\else \,$\mu$K$_{\rm RJ}$\,s$^{1/2}$\fi}
\def\muKHz{\ifmmode \,\mu$K\,Hz$^{-1/2}\else \,$\mu$K\,Hz$^{-1/2}$\fi}
\def\MJysr{\ifmmode \,$MJy\,sr\mo$\else \,MJy\,sr\mo\fi}
\def\MJysrmK{\ifmmode \,$MJy\,sr\mo$\,mK$_{\rm CMB}\mo\else \,MJy\,sr\mo\,mK$_{\rm CMB}\mo$\fi}
\def\microns{\ifmmode \,\mu$m$\else \,$\mu$m\fi}

\def\muK{\ifmmode \,\mu$K$\else \,$\mu$\hbox{K}\fi}
\def\microK{\ifmmode \,\mu$K$\else \,$\mu$\hbox{K}\fi}
\def\muW{\ifmmode \,\mu$W$\else \,$\mu$\hbox{W}\fi}
\def\kms{\ifmmode $\,km\,s$^{-1}\else \,km\,s$^{-1}$\fi}
\def\kmsMpc{\ifmmode $\,\kms\,Mpc\mo$\else \,\kms\,Mpc\mo\fi}
\providecommand{\sorthelp}[1]{}

\def\NHUNIT{\ifmmode {\rm \,cm^{-2}} \else $\rm \,cm^{-2}$ \fi} 

\def\aap{{A\&A}}

\def\apjs{{ApJ Supp.}}
\def\muKcmb{\ifmmode \,\mu$K$_{\rm CMB}$\else \,$\mu$K$_{\rm CMB}$\fi}

\newcommand{\lensing}{lensing}

\newcommand{\OmegaM}{\ifmmode\Omega_{\rm M}\else $\Omega_{\rm M}$\fi}

\newcommand{\commander}{{\tt Commander}}
\newcommand{\lowltt}{{\tt lowl.TT}}

\providecommand{\Planck}{\textit{Planck}}

\providecommand{\text}[1]{\rm{#1}}

\providecommand{\muK}{\mu\rm{K}}

\providecommand{\HEALpix}{{\tt HEALPix}}

\newcommand{\begm}{\begin{pmatrix}}
\newcommand{\enm}{\end{pmatrix}}

\def\pmb#1{\setbox0=\hbox{#1}%
    \kern-.025em\copy0\kern-\wd0
    \kern.05em\copy0\kern-\wd0
    \kern-.025em\raise.0433em\box0}

\def\p2Y{\;_2Y}
\def\m2Y{\;_{-2}Y}
\def\beglet{
  \addtocounter{equation}{1}%
  \setcounter{parentequation}{\value{equation}}%
  \setcounter{equation}{0}%
  \def\theequation{\arabic{parentequation}\alph{equation}}%
  \ignorespaces
}
\def\endlet{
  \setcounter{equation}{\value{parentequation}}%
  \def\theequation{\arabic{equation}}%
}
\providecommand{\beglet}{\begin{subequations}}
\providecommand{\endlet}{\end{subequations}}

\newcommand{\mksym}[1]{\ifmmode {\rm #1}\else #1\fi}

\newcommand{\lowEStwo}{\mksym{{\rm lowE-S2}}}

\providecommand{\text}[1]{\rm{#1}}

\providecommand{\muK}{\mu\rm{K}}

\newcommand\ba{\begin{eqnarray}}
\newcommand\ea{\end{eqnarray}}
\newcommand\bea{\begin{eqnarray}}
\newcommand\eea{\end{eqnarray}}

\newcommand\be{\begin{equation}}
\newcommand\ee{\end{equation}}

\newcommand{\HFI}{{HFI}\xspace} 
\newcommand{\LFI}{{LFI}\xspace}

\newcommand{\srolltwo}{{\tt SRoll2}}

\newcommand{\ELiCAfull}{\texttt{ELiCA.full}}
\newcommand{\ELiCAcross}{\texttt{ELiCA.cross}}
\newcommand{\ELiCAhybrid}{\texttt{ELiCA.hybrid}}
\newcommand{\CamSpec}{\texttt{CamSpec}}
\newcommand{\elica}{\texttt{ELiCA}}
\newcommand{\Plik}{\texttt{Plik}}
\newcommand{\Hillipop}{\texttt{HiLLiPoP}}
\renewcommand{\lensing}{\texttt{Lensing}}
\newcommand{\BAO}{\texttt{BAO}}
\newcommand{\ACT}{ACT}

\newcommand{\ACTplanckcut}{\texttt{ACT.Planckcut}}

\begin{document}

\title{Cross-spectra likelihood for robust $\tau$ constraints from all satellite polarisation data}

\author{
V. Genesini\orcid{0009-0002-5413-9406}\inst{1,2}~\thanks{\email{valentina.genesini@unife.it}}
\and
G. Galloni\orcid{0000-0002-2412-8311}\inst{1,2}~\thanks{\email{giacomo.galloni@unife.it}}
\and
L. Pagano\orcid{0000-0003-1820-5998}\inst{1,2,3}~\thanks{\email{luca.pagano@unife.it}}
\and
P. Campeti\orcid{0000-0002-5637-519X}\inst{2}~\thanks{\email{pcampeti@fe.infn.it}}
\and
M. Lattanzi\orcid{0000-0003-1059-2532}\inst{2}~\thanks{\email{lattanzi@fe.infn.it}}
}

\institute{
\small Dipartimento di Fisica e Scienze della Terra, Universit\`a degli Studi di Ferrara, via Saragat 1, I-44122 Ferrara, Italy\goodbreak
\and
Istituto Nazionale di Fisica Nucleare, Sezione di Ferrara, via Saragat 1, I-44122 Ferrara, Italy\goodbreak
\and
Institut d'Astrophysique Spatiale, CNRS, Univ. Paris-Sud, Universit\'{e} Paris-Saclay, B\^{a}t. 121, 91405 Orsay cedex, France}

\date{\vglue -1.5mm \today \vglue -5mm}

\abstract{
The Thomson scattering optical depth to reionisation, $\tau$, one of the six parameters of the $\Lambda$CDM model, is primarily constrained by the large-scale E-mode polarisation of the Cosmic Microwave Background (CMB). In this work, we present the E-mode Likelihood for Cross-Analysis (\elica), a multi-frequency, harmonic-space likelihood that combines all currently available large-scale satellite polarisation data, namely the \Planck\ \LFI\ 70 GHz channel, the \Planck\ \HFI\ 100 and 143 GHz channels processed with the \srolltwo\ map-making algorithm, and the WMAP Ka, Q, and V bands. The likelihood is built on an extension of the Hamimeche-Lewis formalism to multi-field partial-sky observations. We validate the pipeline using 500 realistic simulations and find that retaining all cross-spectra and the WMAP-LFI auto-spectrum eliminates the significant bias present when all spectra are retained, while preserving comparable uncertainties in the recovered value of $\tau$. From the low-$\ell$ E-mode power spectrum alone, we obtain $\tau = 0.0575_{-0.0058}^{+0.0048}$ (68\% CL). Combining \elica\ with the \Planck\ low-$\ell$ temperature likelihood and the \CamSpec\ high-$\ell$ likelihood, we find $\tau = 0.0581_{-0.0059}^{+0.0048}$ and $\ln(10^{10}A_{\mathrm{s}}) = 3.048_{-0.012}^{+0.011}$. Including \ACT{} DR6 + \Planck\ CMB lensing and DESI DR2 BAO measurements, we derive an upper bound on the total neutrino mass of $\sum m_\nu < 0.069$ eV (95\% CL). Our results, obtained through careful cross-validation of all available large-scale polarisation datasets, robustly confirm that the optical depth remains relatively low. This severely constrains the possibility of explaining, or even significantly reducing, the tension between DESI-BAO and CMB observations with a high value of $\tau$. The \elica\ likelihood is publicly available.
}


\authorrunning{Genesini et al.}

\titlerunning{Cross-spectra likelihood for robust $\tau$ constraints from all satellite polarisation data}

\maketitle
\nolinenumbers
\section{Introduction}\label{sec:intro}

During recombination ($z \sim 1100$), electrons and protons combine to form neutral hydrogen, leading to the decoupling of photons from baryons and to the release of the cosmic microwave background (CMB) radiation. Following recombination, the Universe enters the so-called “Dark Ages”. Whereas dark matter perturbations had already undergone gravitational growth, baryonic perturbations could grow only after decoupling, evolving within the underlying dark matter potential and leading to the formation of structures.
The first stars, galaxies, and quasars emit ultraviolet radiation that reionises the intergalactic medium and releases free electrons, marking the beginning of the Epoch of Reionisation (EoR). Although the exact onset of reionisation remains uncertain, the absence of a Gunn-Peterson trough \cite{Gunn:1965hd} in the spectra of distant quasars at $z \sim 6$ indicates that the Universe was fully ionised by that time. The free electrons released during reionisation leave a measurable imprint on the CMB, modifying the primordial anisotropies through Thomson scattering. This effect is quantified by the Thomson-scattering optical depth to reionisation, $\tau$, i.e. the line-of-sight integral of the free-electron density weighted by the Thomson cross section, which gives the probability that a CMB photon was scattered by free electrons after recombination. At small angular scales, re-scattering suppresses the primordial scalar perturbations by a factor $e^{-2\tau}$, introducing a degeneracy with $A_{\mathrm{s}}$, which sets the amplitude of scalar perturbations.
This degeneracy can be broken through precise measurements of the large-scale E-mode polarisation, as the re-scattering generates a characteristic bump in the power spectrum at multipoles $\ell \lesssim 10$, which is primarily sensitive to $\tau$. For this reason, large-scale CMB observations from satellites are crucial for constraining the reionisation history of the Universe and improving our understanding of the link between the EoR and the formation of the earliest cosmic structures. The first satellite-based constraint came from the Wilkinson Microwave Anisotropy Probe (WMAP)\citep{bennett2012}, whose final 9-year analysis yielded $\tau = 0.089 \pm 0.014$ \citep{Hinshaw2013}. This estimate relied on a dust template constructed from a tracer combining the Finkbeiner–Davis–Schlegel dust intensity model \citep{Finkbeiner:1999aq} with starlight polarisation maps.

Later, the \Planck\ satellite \cite{planck2016-l01} delivered measurements with substantially improved sensitivity and precision on reionisation scales. An important advance was that \Planck\ enabled a move from model-based dust templates to a data-driven treatment of Galactic dust, using the High Frequency Instrument (HFI) 353 GHz polarisation channel as a direct tracer of dust emission. Using the \Planck\ 70 GHz channel of the Low Frequency Instrument (LFI), the 2018 legacy release reported $\tau = 0.063 \pm 0.020$ \citep{planck2016-l05}. The first joint analysis combining LFI and WMAP data was presented by \citet{Lattanzi:2016dzq}, yielding $\tau = 0.066^{+0.012}_{-0.013}$.
Building on this approach, \citet{Natale:2020owc} combined the legacy frequency maps released by the collaborations and performed a pixel-based, low-resolution analysis. By combining their maps with the \Planck\ legacy large-scale temperature solution, they obtained $\tau = 0.069^{+0.011}_{-0.012}$ from a joint temperature–polarisation analysis. In parallel, within the \texttt{BeyondPlanck} framework, \citet{Paradiso:2022fky} used a \commander-based WMAP+LFI combined analysis to measure $\tau = 0.066 \pm 0.013$.
Meanwhile, analyses relying on the \Planck\ HFI, optimised for measuring large-scale polarisation, were limited by significant systematic residuals. These uncertainties prevented the development of a reliable pixel-based likelihood, as the noise covariance matrices of these channels could not be robustly characterised. To overcome these limitations, a harmonic-space likelihood was adopted based on the cross-spectrum between the channels at $100$ and $143$ GHz, yielding $\tau = 0.051 \pm 0.009$ \citep{planck2016-l05}.
A later reanalysis by \citet{Pagano:2019tci} used \srolltwo\ \citep{Delouis:2019bub}, an improved version of the map-making pipeline developed for the \Planck\ legacy release \citep{planck2016-l03}. The resulting maps exhibit reduced large-scale residuals and diminished dipole-distortion systematics, enabling a more precise determination of the optical depth, $\tau = 0.0566^{+0.0053}_{-0.0062}$. Compatible results are obtained, on the same dataset, using an alternative likelihood approach presented in \cite{deBelsunce:2021mec}. Subsequent reanalyses based on the \texttt{NPIPE} processing \citep{planck2020-LVII} find consistent results, $\tau = 0.058\pm0.006$, using cross-spectra from \commander\ CMB maps produced by component separation on frequency maps built from independent data splits \citep{Tristram:2023haj}. More recently, the ground-based CLASS experiment \citep{Eimer:2023esh} has provided an independent large-scale polarisation measurement; in combination with \Planck\ and WMAP data \cite{CLASS:2025khf} reports $\tau = 0.053^{+0.018}_{-0.019}$, consistent with the satellite-based determinations.
Complementarily, \cite{kageura2026} presented a determination of $\tau$ independent of the large-scale CMB polarisation using direct measurements of the neutral hydrogen fraction, finding 
$\tau = 0.0552^{+0.0075}_{-0.0049}$.

To date, no analysis has combined HFI, LFI, and WMAP legacy polarisation data. In the absence of new satellite data, combining these measurements would enable us to fully leverage the available large-scale information to estimate $\tau$ and explore the parameters of the $\Lambda$CDM model and its extensions. To fill this gap, we present the E-mode Likelihood for Cross-Analysis (\texttt{ELiCA}), a harmonic-space likelihood specifically designed to combine the statistical information from the available datasets, based on the approach of \cite{galloni2025}.

The \texttt{ELiCA} estimate therefore serves as an independent cross-check of the CMB-inferred value of $\tau$, also in view of the observations discussed below.
Recent James Webb Space Telescope observations have revealed unexpectedly massive and evolved galaxies at $z > 10$, which, if efficient producers of ionising radiation, could imply an earlier onset of reionisation and thus favour higher values of $\tau$ than those inferred from CMB measurements \citep{munoz2024}.

The recent DESI DR2 analyses \citep{DESI:2025zpo, DESI:2025zgx} have revealed a mild tension ($2.3\sigma$), within the $\Lambda$CDM framework, in the measurements of the matter density fraction, $\Omega_m$, and of the product $H_0 r_d$ of the sound horizon at the drag epoch with the Hubble constant, compared to the constraints from CMB. In extended models where the sum of neutrino masses $\sum m_\nu$ is allowed to vary, this tension is reflected in constraints on this parameter that are tighter than expected, producing in turn a tension with the physical lower bound on the total neutrino mass, $\sum m_\nu \geq 0.06\ \mathrm{eV}$, implied by flavour oscillation experiments. Moreover, if $\sum m_\nu$ is treated as an effective parameter, and allowed to explore the unphysical region of negative masses, the combination of DESI and CMB data shows a preference for $\sum m_\nu<0$. This is due to the positive correlation between $\sum m_\nu$ and $\Omega_m$ in CMB data, which allows the CMB value of $\Omega_m$ to move towards the smaller DESI value by diminishing the neutrino mass. The $\Lambda$CDM tension, as well as the odd preference for effectively negative neutrino masses, can potentially be alleviated by considering more flexible models of the cosmic expansion history, such as those involving dynamical dark energy equation of state ($w_0,w_a$) \citep{Lodha2025}. Some analyses have instead suggested that, due to the negative correlation between $\tau$ and $\Omega_m$, a higher value of $\tau \sim 0.09$ can yield a CMB-preferred $\Omega_m$ to reconcile with the DESI BAO expansion history without invoking evolving dark energy \citep{Sailer_2026}. The apparent preference for sub-minimal neutrino masses would also be alleviated in this way, restoring consistency with neutrino oscillation experiments. As detailed above, such a high value is, at least at face value, in contradiction with all available observations of the large-scale polarisation. Barring non-trivial modifications to the reionisation history, a value of $\tau \sim 0.09$ could only stem from an unaccounted systematic common to all observations of the large-scale polarisation. Another possibility is to consider modifications to the reionisation history that can increase the integrated optical depth while leaving the large-scale CMB polarisation largely unaffected \citep{tan2025}. In this respect, recent works have also suggested that larger values of $\tau$ may arise in scenarios that include an additional early ionisation phase at very high redshift ($z \gtrsim 20$).  Such scenarios differ from the standard late reionisation models commonly adopted in CMB analyses, including the one presented here, and might allow for reconciling a large value of $\tau$ with large-scale CMB polarisation observations without invoking unspecified systematic effects.

The results presented in this work confirm that high $\tau$ values remain in strong tension with large-scale CMB polarisation measurements, at least in the framework of the standard late reionisation model.

The structure of the paper is as follows: in Section \ref{sec:data}, we introduce the datasets and simulations used in this work. Section \ref{sec:likelihood} presents the likelihood implementation and the consistency tests. Section \ref{sec:elica_alone} presents the improved constraint on the reionisation optical depth from the low-$\ell$ E-mode power spectrum alone. Section \ref{sec:cosmology} shows the constraints on $\tau$ and examines the impact of this new likelihood on extensions of the $\mathrm{\Lambda CDM}$ model, focusing in particular on the neutrino sector. 

\section{Data and simulations}\label{sec:data}

To exploit all available information on large-scale E-mode polarisation, we construct a dataset that combines measurements from \Planck\ and WMAP. We use the \Planck\ 100 GHz and 143 GHz channels (hereafter referred to as 100 and 143, respectively) processed with the \srolltwo\ map-making algorithm \citep{Delouis:2019bub}. The level of residual systematics at the largest scales, important for estimating $\tau$, is reduced below the noise level, leaving the cosmic variance as the main source of uncertainty for these channels.
We also include the dataset from \cite{Natale:2020owc} (hereafter WL), which combines the \Planck\ LFI 70 GHz channel with the WMAP Ka, Q, and V bands. Maps are cleaned from synchrotron and dust foreground contamination via template fitting, see \cite{Pagano:2019tci} and \cite{Natale:2020owc} for details. The channels used as foreground tracers, the scaling coefficients, and sky fraction used in the cleaning are reported in Tab.~\ref{tab:coefficients_foreground}. These coefficients are included for completeness and refer to the foreground-cleaned maps used in this analysis. For further details, we point to the original works.

\begin{table}[htbp!]
\centering
\caption{Scaling coefficients for synchrotron (second column) and dust (fourth column) tracers used in
\cite{Natale:2020owc} and \cite{Pagano:2019tci}. The table also reports the sky fraction used in the foreground cleaning.}
\label{tab:coefficients_foreground}

\resizebox{\columnwidth}{!}{
\begin{tabular}{cccccc}
\hline\hline
Band & Synchrotron & $10^{2}\times\alpha$ & Dust & $10^{2}\times\beta$ & Mask \\
\hline
Ka band  & K band  & $32.15\pm0.39$ & 353 GHz & $0.346\pm0.061$ & 55\% \\
Q band   & K band  & $16.51\pm0.39$ & 353 GHz & $0.369\pm0.063$ & 55\% \\
V band   & K band  & $5.27\pm0.27$  & 353 GHz & $0.744\pm0.043$ & 75\% \\
70 GHz   & 30 GHz  & $6.41\pm0.46$  & 353 GHz & $0.966\pm0.041$ & 60\% \\
100 GHz  & K band  & $0.95\pm0.07$  & 353 GHz & $1.86\pm0.015$ & 70\% \\
143 GHz  & Ka band & $1.63\pm0.21$  & 353 GHz & $3.94\pm0.014$ & 70\% \\
\hline
\end{tabular}
}
\vspace{0.5mm}

{\footnotesize \emph{Note.} We correct here a typo in Tab. 1 of \cite{Pagano:2019tci}.}

\end{table}

Following the approach of the \Planck\ Collaboration and given that a robust pixel-based noise covariance cannot be built for the available HFI channels, we implement a suitable multi-frequency harmonic-based likelihood. We use the cleaned maps to compute the power spectra with the Quadratic Maximum Likelihood (QML) estimator \citep{Tegmark:1996qt, Tegmark:2001zv} (see App.~\ref{sec:QML}). In particular, we adopt the publicly available \texttt{pse\_qml} algorithm \footnote{The codes adopted for the component separation and power spectrum estimation are available on the INFN \texttt{GitLab} at \href{https://baltig.infn.it/cosmology_ferrara/lowell-likelihood-analysis}{https://baltig.infn.it/cosmology\_ferrara/lowell-likelihood-analysis}.}.

\begin{figure*}[htbp]
    \includegraphics[width=0.9\textwidth]{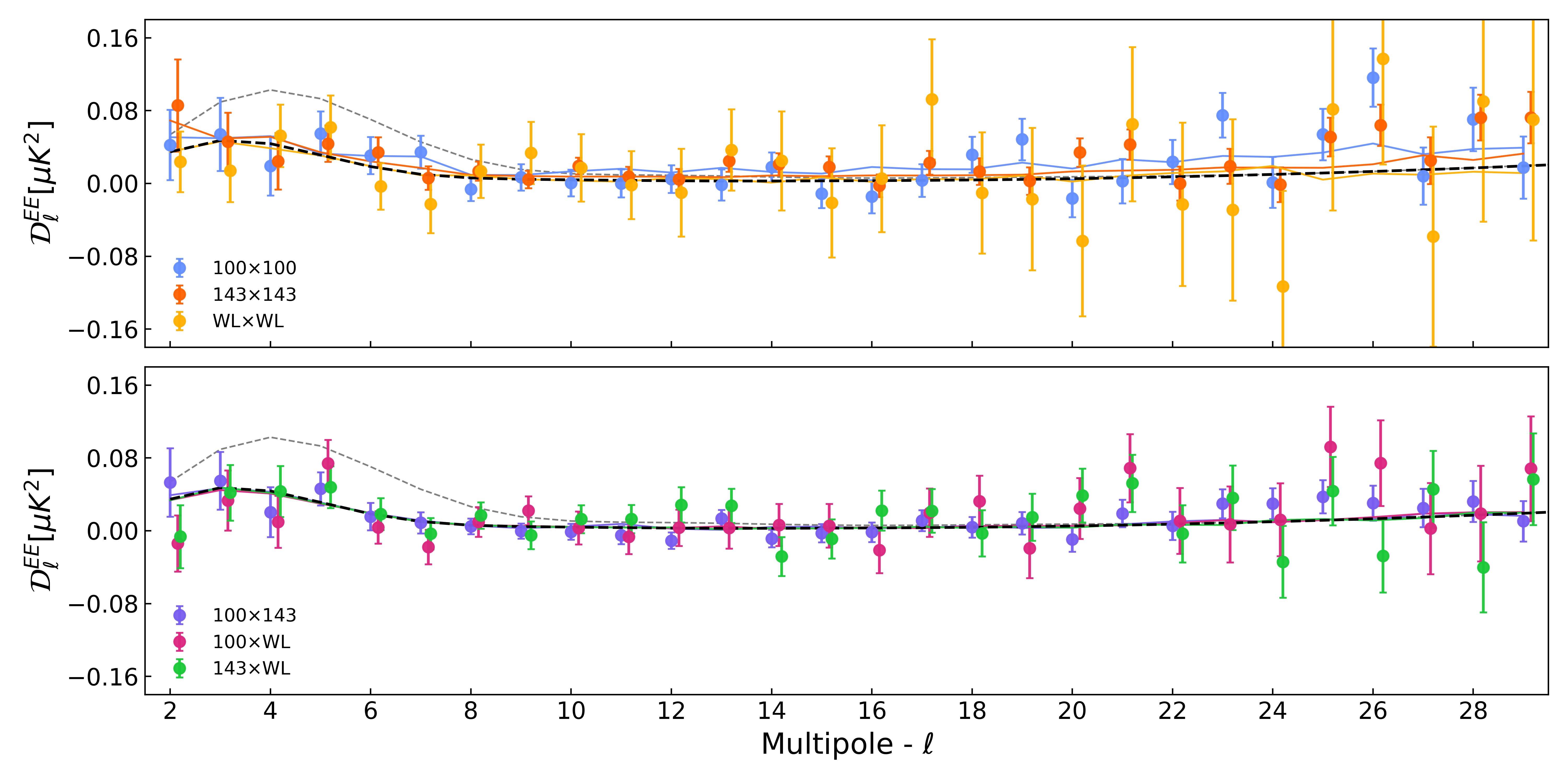}
    \caption{Comparison of data from different channels with the fiducial model. Auto-spectra are shown in the top panel and cross-spectra in the bottom panel. For reference, we also plot the mean spectra obtained from 500 simulations. The error bars represent the standard deviation derived from the same simulations. The fiducial model assumes $\tau = 0.06$ and $A_{\mathrm{s}} = 2.12 \times 10^{-9}$. The grey line corresponds to an alternative model with $\tau = 0.09$ and a rescaled amplitude $A_{\mathrm{s}} = 2.25 \times 10^{-9}$.}
\label{fig:cmb_data} 
\end{figure*}

The Stokes maps are first smoothed in harmonic space using a cosine window function and then downgraded to a \HEALpix\ \citep{Gorski2005} resolution of $N_{\mathrm{side}}=16$. This procedure preserves the signal at low multipoles while preventing small-scale features from aliasing into large scales and gradually suppressing them up to $\ell=48$. Only pixels outside the Galactic masks are included in the analysis to minimise foreground contamination. For 100 and 143, we adopt a mask retaining 50\% of the sky, while for WL we retain 54.2\%. The corresponding sky fractions are reported in Tab.~\ref{tab:PTE}. The QML algorithm requires accurate noise covariance matrices to provide unbiased estimates of the power spectra. For this reason, we use the noise covariance matrices from \citet{Natale:2020owc} for the WL dataset, and the FFP8 covariance matrices \citep{keskitalo2010} for the \Planck\ HFI maps, following the \Planck\ Collaboration approach. The resulting unbiased E-mode spectra for each frequency channel are shown in Fig.~\ref{fig:cmb_data}.

To validate the performance and robustness of the harmonic-space likelihood, we generate 500 realisations of full-sky CMB polarisation maps based on a fiducial $\Lambda$CDM cosmology. The optical depth is fixed at $\tau = 0.06$, while the scalar amplitude $A_{\mathrm{s}}$ is adjusted such that $10^9 A_{\mathrm{s}} e^{-2\tau}= 1.884$. All the remaining cosmological parameters are taken equal to their maximum a posteriori estimates inferred from the combined \lowEStwo, \lowltt\ and high-$\ell$ \texttt{Plik} 2018 datasets, as presented in Table 3, column 4, of \citet{Pagano:2019tci}. Each simulated map is processed identically to the data: convolved with the instrumental beam, smoothed in harmonic space using a cosine window function, and downgraded to a \HEALpix\ resolution of ${\mathrm{N_{side}}}=16$. Realistic mock datasets are generated by adding noise (and systematic effects) realisations to the CMB signal. For \Planck\ HFI, we use simulations described in \citet{Delouis:2019bub}, which include instrumental noise, systematics, and foregrounds. For the WL dataset, independent Gaussian noise realisations are generated from the corresponding covariance matrix \citep{Natale:2020owc,planck2016-l02}. All the simulations are independently passed through the template fitting procedure. As for \srolltwo\ only 500 independent noise realisations are available, which limits the number of statistically independent mock simulations that can be generated for each spectrum. Nevertheless, the ensemble of 500 simulations provides a statistically robust and realistic testbed for validating the harmonic-space likelihood framework. To show the consistency of the spectra, we compute the harmonic $\chi^2$ statistic:

\begin{equation}
    \chi^2 = \sum_{\ell,\ell^\prime}^{\ell_{\text{max}}} \left(C_\ell-C_{\ell}^{\text{th}}\right)^{T} M_{\ell\ell^{\prime}}^{-1} \left(C_\ell^\prime-C_{\ell^\prime}^{\text{th}}\right),
\end{equation}
where $M_{\ell\ell^{\prime}}$ is the covariance matrix estimated from half of the available simulations, with the remaining half used in the test to compute the term $(C_\ell-C_{\ell}^{\text{th}})$.

We report in Tab.~\ref{tab:PTE} the empirical probability of observing a value of $\chi^2$ greater than that for the data at different $\ell_{\text{max}}$. This is the Probability To Exceed (PTE), which quantifies how well the observed data match the theoretical model. The results follow what is shown in \citet{Natale:2020owc} and \citet{Pagano:2019tci}. In particular, the behaviour of the WL $\times$ WL power spectra is consistent with the results presented in \citet{Natale:2020owc} (see Tab.~7). The WL $\times$ HFI cross-spectra, presented here for the first time, exhibit PTE values consistent with the expected distribution of the test statistic, and therefore constitute a solid preliminary validation before proceeding to the likelihood analysis.

\begin{table}[htbp!]
\begingroup
\centering  
\caption{Probability to exceed for the unbiased auto- and cross-spectra used in this work, evaluated for different $\ell_{\rm max}$ values.}
\label{tab:PTE}
\nointerlineskip
\vskip -3mm
\setbox\tablebox=\vbox{
   \newdimen\digitwidth
   \setbox0=\hbox{\rm 0}
   \digitwidth=\wd0
   \catcode`*=\active
   \def*{\kern\digitwidth}

   \newdimen\signwidth
   \setbox0=\hbox{+}
   \signwidth=\wd0
   \catcode`!=\active
   \def!{\kern\signwidth}

\halign{\hbox to 0.6in{#\leaderfil}\tabskip=1em&
  \hfil#\hfil\tabskip=1em&
  \hfil#\hfil\tabskip=1em&
  \hfil#\hfil\tabskip=1em&
  \hfil#\hfil\tabskip=1em&
  \hfil#\hfil\tabskip=0pt\cr
\noalign{\doubleline}
\noalign{\vskip -2pt}
\omit\hfil Channel \hfil & \hfil Mask \hfil & $\ell_{\text{max}}=10$ & $\ell_{\text{max}}=15$ & $\ell_{\text{max}}=29$\cr
\noalign{\vskip 3pt\hrule\vskip 5pt}
$100\times100$ & 50\% & $84\%$ & $98.8\%$ & $69.6\%$\cr
$100\times143$ & 50\% & $98.4\%$ & $74\%$ & $74.4\%$\cr
$100\times \textrm{WL}$ & 44.1\% & $36\%$ & $76.4\%$ & $70.4\%$\cr
$143\times143$ & 50\% & $67.2\%$ & $48\%$ & $8.8\%$\cr
$143\times\textrm{WL}$ & 44.1\% & $87.2\%$ & $64\%$ & $70.4\%$\cr
$\textrm{WL}\times\textrm{WL}$  & 54.2\% & $87.6\%$ & $97.2\%$ & $98.8\%$\cr
\noalign{\vskip 5pt\hrule\vskip 3pt}}}
\endPlancktable
\endgroup
\end{table}

The PTE values fluctuate at the 2–3$\sigma$ level across individual realisations, indicating no significant deviations and confirming that the spectra can be reliably used in the subsequent likelihood analysis. Further validations of the framework are provided in Section \ref{sec:likelihood}.

\section{Likelihood implementation and validation}\label{sec:likelihood}
In this section, we present the harmonic-based likelihood algorithm implemented in this analysis. The methodology follows \cite{galloni2025}, who extends the Hamimeche-Lewis (HL) formalism \citep{Hamimeche_2008} to properly account for multi-field partial-sky coverage and uncertainties in the noise bias at large angular scales. The HL formalism was originally introduced to provide a likelihood capable of accounting for the non-Gaussian distribution of power-spectrum estimators derived from correlated Gaussian fields observed on a partial sky. It can be adapted to any multi-field scenario, provided that the correlated fields considered are Gaussian and statistically isotropic. 
Since the 100, 143, and WL datasets satisfy these conditions, we adopt this methodology to construct a harmonic-based likelihood for large-scale E-mode polarisation.
The HL approach involves rewriting the exact full-sky likelihood, derived under the assumption of Gaussian CMB perturbations, as:

\begin{equation}\label{eq:HL_def}
-2 \ln \mathcal{L}\!\left(\hat{\mathbf{C}}_{\ell} \mid \mathbf{C}_\ell\right) 
= \sum_{\ell \ell'} X^{T}_{\ell} \,M^{-1}_{\ell \ell'}\big|_{\mathrm{fid}} \,X_{\ell'},
\end{equation}
where $\hat{\mathbf{C}}_{\ell}$ is the estimated power spectrum of the true underlying power spectra $\mathbf{C}_\ell$ and $M^{-1}_{\ell \ell'}$ is the precision matrix used to quantify correlations among different multipoles, estimated from simulations evaluated for a fiducial model. The HL likelihood approximation has a structure of a Gaussian distribution in

$X_\ell$. In this analysis, $X_\ell$ is given by:
\begin{equation}
    X_\ell=(X^{100\times100}_\ell, X^{100\times143}_\ell, X^{100\times \textrm{WL}}_\ell, X^{143\times143}_\ell, X^{143\times \textrm{WL}}_\ell, X^{\textrm{WL}\times \textrm{WL}}_\ell) .
\end{equation}
This vector is built as $\mathbf{X_\ell} = \mathrm{vecp}(\mathbf{C}_{\ell})$, i.e, we retain only the independent elements of the matrix:
\begin{equation}
    \mathbf{C}_{g,\ell} = \mathbf{C}_{f,\ell}^{1/2}\,
    \mathbf{U}_{\ell}\, \mathbf{g}(\mathbf{D}_{\ell})\, \mathbf{U}^{T}\,
    \mathbf{C}_{f,\ell}^{1/2} \, .
    \label{eq:Cg_definition}
\end{equation}

Here, $\mathbf{C}_{f,\ell}$ are the fiducial power spectra, $\mathbf{U}_{\ell}$ is the matrix that diagonalises $\mathbf{C}_{\ell}^{-1/2} \, \hat{\mathbf{C}}_{\ell} \, \mathbf{C}_{\ell}^{-1/2}$ and the diagonal matrix $\mathbf{D}_{\ell}$ contains the corresponding eigenvalues. The function $\mathbf{g}( \mathbf{D}_{\ell})$ acts element wise on the diagonal of $\mathbf{D}_{\ell}$ and is defined as \citep{Hamimeche_2008}: 
\begin{equation}
\begin{aligned}
    \qty[\mathbf{g}(\mathbf{D}_{\ell})]_{ij}
    &= g(D^{ii}_\ell) \delta_{ij} \\
    &= \operatorname{sign}(D^{ii}_\ell - 1) \sqrt{
      2 \bigl( D^{ii}_\ell - \ln D^{ii}_\ell - 1 \bigr)}\, 
      \delta_{ij} \, .
\end{aligned}
\label{eq:g_definition}
\end{equation}
In our multi-frequency approach, the $\mathbf{C}_\ell$ takes the form:

\begin{equation}
\label{eq:Cl_matrix}
\begin{aligned}
&\mathbf{C}_\ell =\begin{bmatrix*}[l]
C_\ell^{100\times 100} + N_\ell^{100} & C_\ell^{100\times 143} & C_\ell^{100\times \textrm{WL}} \\[1pt]
C_\ell^{100\times 143} & C_\ell^{143\times 143} + N_\ell^{143} & C_\ell^{143\times \textrm{WL}} \\[1pt]
C_\ell^{100\times \textrm{WL}} & C_\ell^{\textrm{WL}\times 143} & C_\ell^{\textrm{WL}\times \textrm{WL}} + N_\ell^{\textrm{WL}}
\end{bmatrix*}, 
\end{aligned}
\end{equation}
where we assume that the noise is uncorrelated between different channels. Under this assumption, an accurate estimate of the noise bias is required only for the auto-spectra, as prescribed by \citet{Hamimeche_2008}. While some analyses address this issue by relying only on a single cross-spectrum \citep{Mangilli_2015} or on concatenations of a few of them \citep{vanneste:tel-02426412}, the method of \cite{galloni2025} allows retaining all well-characterised auto-spectra, marginalising over those with poorly determined noise bias. This is possible because the HL likelihood preserves the structure of a Gaussian likelihood in $X_\ell$, thus enabling the use of the standard marginalisation rule. In our analysis, this framework is particularly relevant as \citet{Natale:2020owc} provides a reliable noise covariance matrix for the WL dataset, allowing us to include its auto-spectrum in the computation. Since no such estimate exists for the HFI channels, their auto-spectra must be marginalised over. This marginalisation is performed on the vector $X_{\ell}$; specifically, we integrate out the components $X^{100\times100}_\ell$ and $X^{143\times143}_\ell$. The vector used in the analysis, therefore, becomes: 
\begin{equation}
    X_\ell=(X^{100\times143}_\ell, X^{100\times \textrm{WL}}_\ell, X^{143\times \textrm{WL}}_\ell, X^{\textrm{WL}\times \textrm{WL}}_\ell) .
\end{equation}
The covariance matrix is adjusted in such a way that the elements associated with $C_\ell^{100\times 100}$ and $C_\ell^{143\times 143}$ are removed.

A further complication arises from the incomplete sky coverage. Although auto-spectra are positive definite on the full-sky, masking induces the matrix $\mathbf{C}_{\ell}^{-1/2}\hat{\mathbf{C}}_{\ell}\mathbf{C}_{\ell}^{-1/2}$ to develop negative eigenvalues. Thus, we add an offset $O_\ell$ for each auto-spectra to mitigate this cut-sky effect and stabilise the eigenvalue problem. In particular, the offset is chosen such that more than 99\% of realisations yield a positive-definite transformed matrix, which is consistent with what is described in \cite{Mangilli_2015}.

Moreover, following the procedure introduced in \cite{Mangilli_2015}, Eq.~\ref{eq:g_definition} is also modified as follows:
\begin{equation}
g(x) \longrightarrow g(|x|)\, \mathrm{sign}(x)
\end{equation}

While this modification has minimal effect once the offset is applied, it should be emphasised that its purpose is strictly to ensure numerical stability rather than to address statistical considerations.

As a final methodological remark, we note that when inferring parameters from a Gaussian-distributed dataset using likelihood evaluation, one requires a covariance matrix to describe the data errors and their correlations. If this matrix is not known a priori, it must be estimated from mock realisations and thus becomes a random quantity, with its own uncertainty. In practice, however, we have access to only a limited number of simulations, which introduces uncertainty into the estimated covariance matrix. While the estimated covariance matrix is an unbiased estimator of the true covariance, the estimated precision matrix is a biased estimator of the true inverse covariance matrix. To properly account for this, we adopt the Sellentin–Heavens (SH) approximation (see \cite{Sellentin_2015}), which marginalises over the true covariance matrix conditioned on its estimated value.

The resulting form of our likelihood is thus:
\begin{equation}
-2\ln \mathcal{L}(\hat{\mathbf{C}}_\ell \mid \mathbf{C}_\ell) 
= N \ln \left[ 1 + \frac{ X_\ell^\mathrm{T} \, M^{-1}_{\ell \ell'}\big|_{\mathrm{fid}} \, X_{\ell'} }{N-1} \right] \,,
\end{equation}
where N quantifies the number of degrees of freedom entering the covariance evaluation.

To validate the likelihood, we first build a uniform grid of 2401 theoretical power spectra, spanning $\tau$ values from 0.01 to 0.16. For each spectrum, the amplitude of scalar perturbations is adjusted to satisfy the condition $A_{\mathrm{s}}e^{-2\tau}=1.884$. At the same time, all other cosmological parameters are fixed to the \lowEStwo, \lowltt\ and high-$\ell$ \Plik\ 2018 datasets, as presented in Table 3 of \citet{Pagano:2019tci}. Following the pipeline for power spectrum estimation introduced in Section \ref{sec:data}, we adopt the fiducial power spectrum corresponding to $\tau=0.06$. To test the likelihood, the 500 estimated CMB power spectra are split into two sets of 250. This partition sets $\mathrm{N}=250$ for the SH correction, consistent with the available degrees of freedom appearing in the covariance matrix estimation. The first set is used to evaluate the covariance matrix and the offset, while the second is used as a test set. The noise bias $N_\ell$ for the auto-spectra, introduced in Eq.~\ref{eq:Cl_matrix}, is estimated by applying the QML estimator to noise-only maps. 
The resulting unbiased noise auto-spectra are then corrected by adding the bias term obtained from the QML estimator applied to the maps containing both the CMB signal and the noise, introduced in Section \ref{sec:data}. More details on the noise bias estimation are provided in App.~\ref{sec:QML}.

To assess which combination of datasets yields the most robust likelihood, we consider three distinct configurations:
\begin{enumerate}
    \item \ELiCAfull: all auto- and cross-spectra are included.
    \item \ELiCAcross: only cross-spectra are included, and all auto-spectra are marginalised over.
    \item \ELiCAhybrid: all cross-spectra and the WL auto-spectrum are included, while the HFI auto-spectra are marginalised over.
\end{enumerate}
We stress that, although HFI auto-spectra are marginalised over in \ELiCAhybrid{} and \ELiCAcross{} at the level of the independent elements of the vector $X_{\ell}$, yet they remain part of the Gaussianisation procedure. 

\begin{figure}[htbp]
    \includegraphics[width=0.50\textwidth]{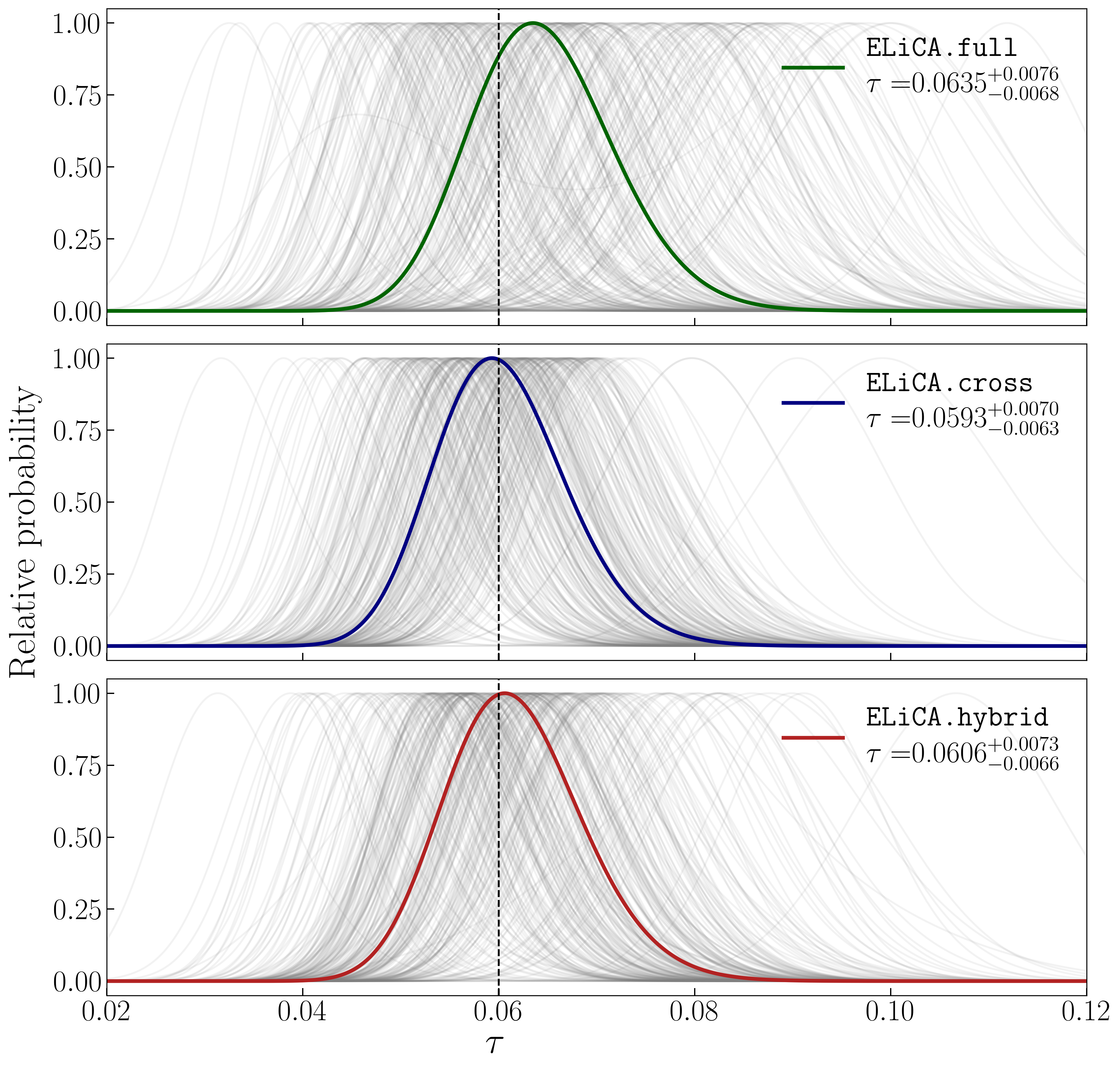}
    \caption{Posterior distributions for the 250 simulated data sets. Each panel corresponds to a different likelihood configuration. Grey curves show the posteriors from individual simulations, while the solid coloured curves correspond to the \ELiCAfull{} (green), \ELiCAhybrid{} (red), \ELiCAcross{} (blue) cases. The vertical dashed line marks the reference value $\tau = 0.06$.}
\label{fig:validation}
\end{figure}

In Fig.~\ref{fig:validation}, we present the posterior distributions from 250 simulated data sets, used to validate the performance of the different likelihood configurations, and for each case, we quantify the deviation from the theoretical value in units of the standard error of the mean, $\hat{\sigma}$. For \ELiCAfull{}, we observe a large deviation of $7.55\,\hat{\sigma}$, as expected due to the inclusion of HFI auto-spectra noise, which is not well characterised. For \ELiCAhybrid{} and \ELiCAcross{}, the corresponding deviations are $1.34\,\hat{\sigma}$ and $-1.63\,\hat{\sigma}$, respectively. The \ELiCAhybrid{} case shows that retaining the WL auto-spectrum, whose noise bias can be reliably reconstructed, allows us to exploit its information without introducing biases in the parameter estimation. Indeed, \ELiCAhybrid{} is the most complete and reliable configuration that can be constructed from the available datasets. 

\texttt{ELiCA} and all the dataset combinations presented above are implemented as a publicly available Python package\footnote{\elica\ likelihood available at \url{https://github.com/Vgenesin/ELICA}}.

\section{Results with \texttt{ELiCA} alone}\label{sec:elica_alone}

\begin{figure}[htbp]
    \centering
    \includegraphics[width=0.45\textwidth]{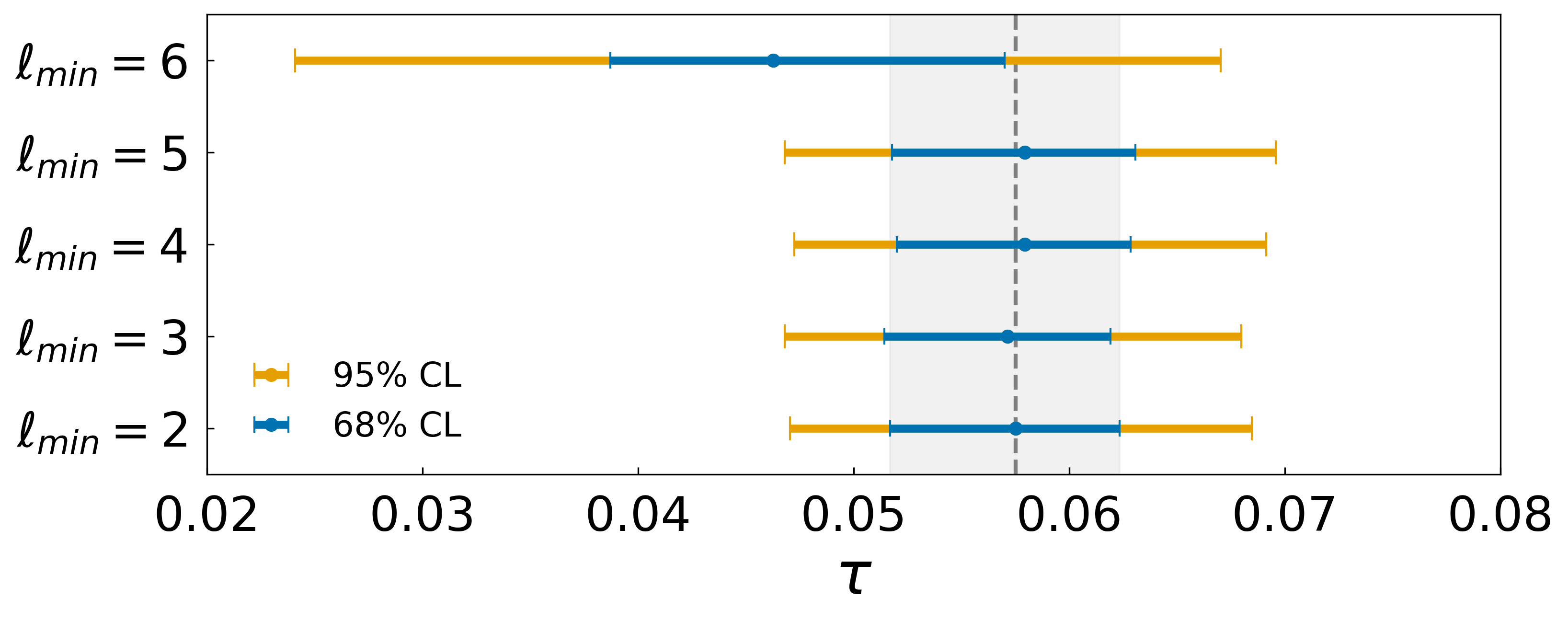}
    \caption{Values of $\tau$ obtained by varying the minimum multipole included in the likelihood analysis. The grey shaded band shows the baseline value defined in Eq.~\ref{eq:reference_point}, with the dashed line indicating the mean and the band width corresponding to the 68\% confidence level.}
\label{fig:data_dif_lmin}
\end{figure}

In Section \ref{sec:likelihood} we validated \texttt{ELiCA}, identifying \ELiCAhybrid{} and \ELiCAcross{} as the most reliable combinations of our datasets. Using the same validation grid, in which the optical depth is fixed at $\tau = 0.06$ and the scalar amplitude $A_{\mathrm{s}}$ is adjusted so that $10^9 A_{\mathrm{s}} e^{-2\tau} = 1.884$, we report the mean value and the 68\% confidence level bounds on the optical depth derived from the data:

\begin{equation} \label{eq:reference_point}
    \tau = 0.0575^{+0.0048}_{-0.0058}
    \quad
    \parbox[c]{4.5cm}{\raggedright 68\%, \ELiCAhybrid}
\end{equation}
and
\begin{equation}
    \tau = 0.0564^{+0.0048}_{-0.0057}
    \quad
    \parbox[c]{4.5cm}{\raggedright 68\%, \ELiCAcross.}
\end{equation}
To further demonstrate the reliability of our implementation of \ELiCAhybrid{}, we repeat this analysis by varying the minimum multipole, $\ell_{\text{min}}$, as shown in Fig.~\ref{fig:data_dif_lmin} 

The figure shows that, even when we remove the multipoles that contain most of the physical information on the EoR, the resulting estimates remain fully compatible. It is worth observing that in all cases, the constraints show that large scales alone are sufficient to rule out higher values of the optical depth.

\begin{figure}[htbp]
    \centering
    \includegraphics[width=0.45\textwidth]{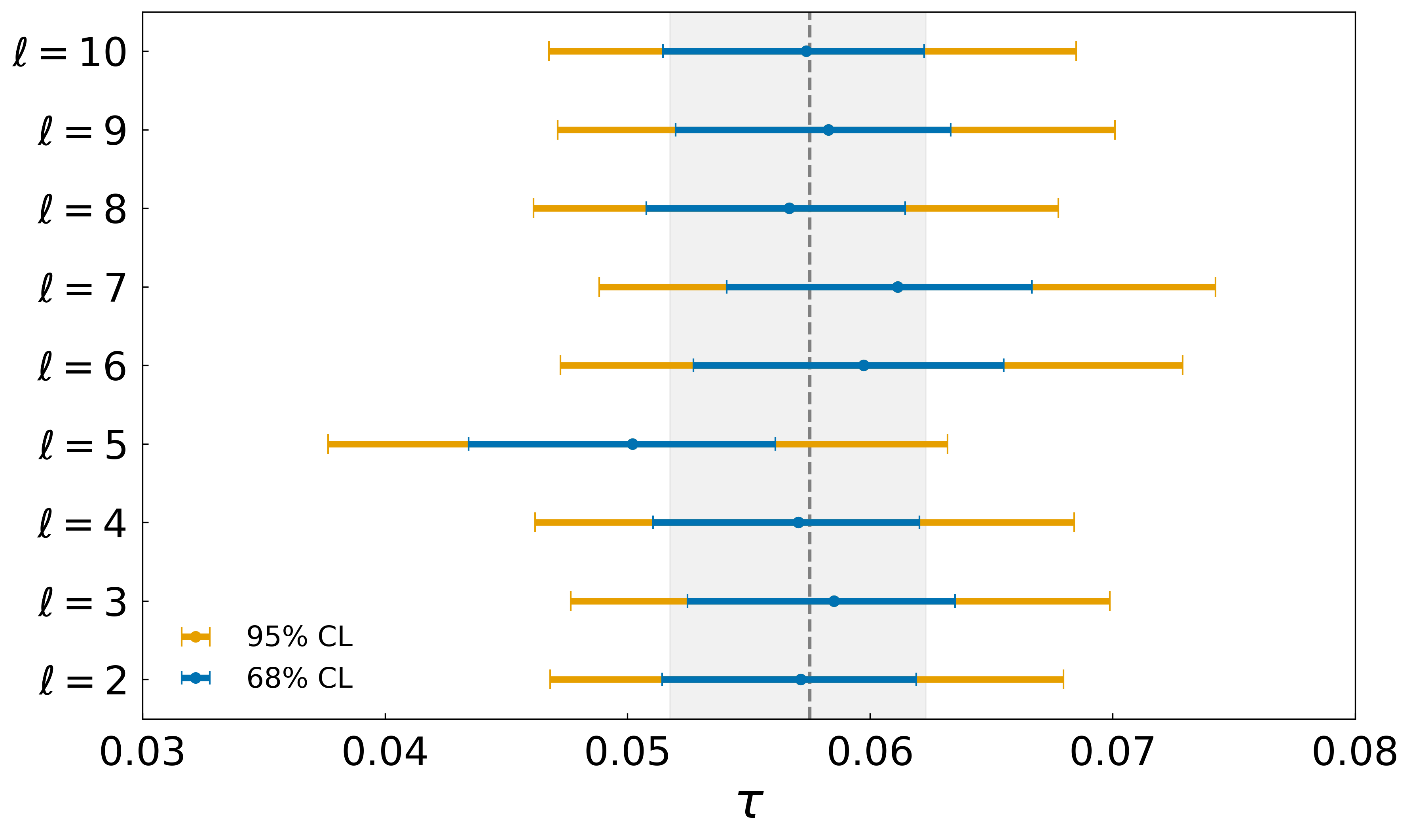}
    \caption{Posterior distributions of $\tau$ obtained by removing one multipole at a time from the likelihood analysis. The grey shaded band shows the baseline value defined in Eq.~\ref{eq:reference_point}, with the dashed line indicating the mean and the band width corresponding to the 68\% confidence level.}
\label{fig:data_removing_l}
\end{figure}

In addition, we perform a stability check by removing one multipole at a time. The results are shown in Fig.~\ref{fig:data_removing_l}. A deviation is observed at $\ell=5$, in agreement with the results of the HFI cross-spectrum analysis presented by \citet{Pagano:2019tci} (see Fig.~14). This behaviour is expected, as the HFI channels retain most of the constraining power. Unlike the aforementioned HFI analysis, no equivalent test has been performed for WL$\times$WL, so a direct comparison is not feasible. 

\section{Impact on cosmology}\label{sec:cosmology}
In this section, we assess the cosmological implications of \texttt{ELiCA} when combined with other cosmological likelihoods.

To explore this, we perform a series of Markov Chain Monte Carlo (MCMC) analyses using the \texttt{Cobaya} framework \citep{Torrado_2021}, with theoretical power spectra computed using \texttt{camb} \citep{Lewis:1999bs, Howlett:2012mh}. The chain convergence is verified through the Gelman-Rubin statistic, with a target threshold of $R-1=0.001$. 

In all analyses, we adopt the default reionisation model of \texttt{camb}, corresponding to the \texttt{TANH} parameterisation, which describes a rapid transition of the intergalactic medium from a nearly neutral to a fully ionised state, exploring alternative parametrisations \citep{Douspis:2015nca, Montero-Camacho:2024dzs} is beyond the scope of this work. We note, however, that \citet{Ilic:2025idl} found that the dependence of current $\tau$ constraints on the assumed reionisation-history parametrisation is small, amounting to about $10\%$ of the present error bar.

Following the methodology adopted by the \Planck\ Collaboration, we first combine \texttt{ELiCA} with the 2018 low-$\ell$ temperature likelihood \cite{planck2016-l05}, here named $\lowltt$, and sample the parameters $\ln(10^{10}A_{\mathrm{s}})$ and $\tau$, while keeping the remaining cosmological parameters fixed to the best-fit values shown in \cite{Pagano:2019tci}. We report the mean value and the 68\% confidence level intervals:

\begin{equation}\label{eq:ELiCA_constraints}
\left.
\begin{aligned}
\tau &= 0.0587_{-0.0061}^{+0.0051}, \\
\ln(10^{10} A_{\mathrm{s}}) &=2.975_{-0.049}^{+0.048},
\end{aligned}
\;
\right\}
\quad
\parbox[c]{3.5cm}{\raggedright 68\%, \lowltt\ \\
+ \ELiCAhybrid{}.}
\end{equation}

The resulting posterior distributions are consistent with previous analyses based on large-scale polarisation data. In particular, \cite{Pagano:2019tci} reported $0.0579^{+0.0056}_{-0.0067}$ from the cross-spectrum of the HFI 100 and 143 GHz channels, while the pixel-based analysis combining WMAP and LFI, with the TE correlation set to zero, found  $0.062\pm0.012$ \citep{Natale:2020owc}. The comparison with \cite{Natale:2020owc} is motivated by the fact that nulling TE effectively ignores TE correlations, using TT and EE only, similar to our approach. Although not identical to our treatment, this provides a useful consistency check. The relatively low value of $\ln(10^{10} A_{\mathrm{s}})$ is also expected when considering large angular scales only. It is consistent with the "low-$\ell$ anomaly" and is driven by a deficit in the measured TT power spectrum at large scales, which pulls down the value of As.

To further test the performance of \texttt{ELiCA}, we extend the analysis to the full set of $\Lambda$CDM parameters. Following the approach of DESI collaboration, we include the high-$\ell$ temperature and polarisation likelihood \CamSpec{} \citep{Efstathiou_2021, Rosenberg_2022}, obtaining: 

\begin{equation}\label{eq:ELiCA_CamSpec_constraints}
\left.
\begin{aligned}
\tau &= 0.0581_{-0.0059}^{+0.0048}, \\
\ln(10^{10} A_{\mathrm{s}}) &= 3.048_{-0.012}^{+0.011},
\end{aligned}
\;
\right\}
\quad
\parbox[c]{3.5cm}{\raggedright 68\%, \lowltt\ \\
+ \ELiCAhybrid{}\\ + \CamSpec{}}
\end{equation} 

In App.~\ref{sec:altre_highl}, we additionally report the constraints obtained by combining \texttt{ELiCA} with alternative high-$\ell$ likelihoods, such as \Plik\ \citep{planck2016-l05} and \Hillipop\ \citep{Tristram:2023haj}, allowing for a direct comparison of the impact of different high-$\ell$ likelihoods on the inferred cosmological parameters. 
In \citet{Pagano:2019tci} and \citet{Natale:2020owc}, the use of the \Plik\ high-$\ell$ likelihood caused an increase in the value of $A_{\mathrm{s}}$ and pulled $\tau$ toward higher values. In contrast, the \CamSpec\ likelihood favours a lower value of $A_{\mathrm{s}}$ \citep{Rosenberg_2022}, mitigating the dragging of $\tau$ when including high-$\ell$ data. 

Having obtained constraints on $\tau$ and $A_{\mathrm{s}}$ using \lowltt\ and \CamSpec, we extend the analysis by including datasets complementary to the CMB temperature and polarisation spectra. Accordingly, we introduce in our pipeline the joint \ACT{} DR6 + \Planck\ CMB lensing likelihood\footnote{We use the likelihood labelled \texttt{actplanck\_baseline} available at \url{https://github.com/ACTCollaboration/act_dr6_lenslike}.}\citep{ACT:2023dou, ACT:2023kun, Carron:2022eyg}, in the following denoted simply as \lensing{}, and the latest BAO distance measurements from DESI DR2, in the following denoted simply as \BAO{} \citep{DESI:2025zpo, DESI:2025zgx}. 

The resulting constraints are:
\begin{equation}\label{eq:ELiCA_CamSpec_lensing_BAO}
\left.
\begin{aligned}
\tau &= 0.0632_{-0.0062}^{+0.0051}, \\
\ln(10^{10} A_{\mathrm{s}}) &= 3.0585_{-0.011}^{+0.0097},
\end{aligned}
\;
\right\}
\quad
\parbox[c]{3.5cm}{\raggedright 68\%, \lowltt \\
+ \ELiCAhybrid{} \\ + \CamSpec\ 
+ \lensing{} + \BAO.}
\end{equation}

This constraint on the amplitude of scalar perturbations corresponds to the following bound on $\sigma_{8}$:
\begin{equation}\label{eq:sig8_ELiCA_CamSpec_lensing_BAO}
\begin{aligned}
\sigma_{8} &= 0.8104^{+0.0042}_{-0.0046}
\end{aligned}
\quad
\parbox[c]{4.5cm}{\raggedright 68\%, \lowltt\ \\
+ \ELiCAhybrid{} + \CamSpec\ \\
+ \lensing{} + \BAO,}
\end{equation}

which sets the amplitude of the matter power spectrum on the scale of $8 h^{-1}$ Mpc.
Assuming the tanh parametrisation of the ionisation fraction, the constraint on the optical depth can be translated into a reionisation mid-point redshift of:
\begin{equation}\label{eq:zre_ELiCA_CamSpec_lensing_BAO}
\begin{aligned}
z_\mathrm{re} &= 8.52^{+0.51}_{-0.58}
\end{aligned}
\quad
\parbox[c]{4.5cm}{\raggedright 68\%, \lowltt\ \\
+ \ELiCAhybrid{} + \CamSpec\ \\
+ \lensing {} + \BAO.}
\end{equation}

In our baseline analysis, we fixed $\sum m_\nu=0.06$ eV, corresponding to the minimum value allowed by flavour oscillation experiments under the normal mass ordering. We model the neutrino mass spectrum with one massive and two massless neutrino states. This approximation is appropriate given the sensitivity of cosmological observations.

Here, we relax this assumption and allow $\sum m_\nu$ to vary, adopting a model with three degenerate massive neutrinos, which provides an effective description of both normal and inverted mass orderings. We further impose a physically motivated uniform prior, $\sum m_\nu>0$ eV, to ensure positive-definite values. This setup allows us to assess how the improved constraints on $\tau$ obtained with \texttt{ELiCA} translate into constraints on the neutrino sector. Thus, the $\mathrm{\Lambda CDM}+\sum m_\nu$ extension at 95\% confidence level gives: 
\begin{equation}\label{eq:mnubound_ELiCA_CamSpec_lensing}
\begin{aligned}
\sum m_\nu &< 0.24
\end{aligned}
\quad
\parbox[c]{4.5cm}{\raggedright 95\%, \lowltt\ \\
+ \ELiCAhybrid{} + \CamSpec\ \\
+ \lensing{}}
\end{equation}
Adding the DESI DR2, we obtain;
\begin{equation}\label{eq:mnubound_ELiCA_CamSpec_lensing_BAO}
\begin{aligned}
\sum m_\nu &< 0.069
\end{aligned}
\quad
\parbox[c]{4.5cm}{\raggedright 95\%, \lowltt\ \\
+ \ELiCAhybrid{} + \CamSpec\ \\
+ \lensing{} + \BAO.}
\end{equation}
This bound is slightly weaker than the one reported by the DESI Collaboration, $\sum m_\nu <0.0642$ \citep{elbers2025, DESI:2025zgx}. This difference can be attributed to a different large-scale E-mode polarisation likelihood, as the DESI analysis relies on \texttt{SimALL} \cite{planck2016-l05}, which favours lower values of the optical depth and induces a shift in the inferred bound on the $\sum m_\nu$.

Finally, we extend the analysis to the $\mathrm{\Lambda CDM+ \mathrm{N_{eff}}}$ model, where $\mathrm{N_{eff}}$ quantifies the effective number of relativistic species. Previously fixed at 3.044, $\mathrm{N_{eff}}$ is now varied with a uniform prior, giving the following constraint:
\begin{equation}\label{eq:Neff_all_likelihoods}
\begin{aligned}
{\mathrm{N_{eff}}} &= 3.21\pm0.17
\end{aligned}
\quad
\parbox[c]{4.5cm}{\raggedright 68\%, \lowltt\ \\
+ \ELiCAhybrid{} + \CamSpec\ \\
+ \lensing{} + \BAO,}
\end{equation}
which is compatible with the one reported in \cite{elbers2025} $3.23\pm0.18$ at 68\% confidence level. 
In this section, we presented the bound obtained following the DESI pipeline, while App.~\ref{sec:ACT_DR6} reports the corresponding constraints derived using the ACT DR6 pipeline, showing consistent results for the same parameters.

\section{Conclusions}\label{sec:conclusions}

In this work, we have introduced \elica{} (E-mode Likelihood for Cross-Analysis), a harmonic-space likelihood specifically designed to jointly exploit all available large-scale CMB polarisation data from \Planck, including both HFI and LFI \citep{Pagano:2019tci}, and WMAP \citep{Hinshaw2013}. This represents the first combined analysis of these datasets and provides the tightest constraint to date on the Thomson scattering optical depth to reionisation, $\tau$. The likelihood is based on the extended Hamimeche–Lewis formalism \citep{Hamimeche_2008}, which naturally accommodates multi-field partial-sky observations and enables the marginalisation of auto-spectra affected by not well characterised noise bias \citep{galloni2025}. 
This is particularly relevant for HFI channels, whose noise properties remain poorly characterised, yet still play a necessary role in the Gaussianisation procedure. 

The implementation of the likelihood further depends on simulations, from which the covariance matrix is estimated. To construct these, noise realisations are added to CMB-only maps, yielding mock maps that are then processed with the QML algorithm to obtain minimum-variance power spectra. In practice, the available noise simulations differ across datasets: HFI is limited to 500 noise realisations from previous \srolltwo\ analyses, whereas simulations for WMAP and LFI can be generated directly from the pixel-based noise covariance matrices.
Given the limited number of independent simulations, we introduce the Sellentin–Heavens correction to account for this \citep{Sellentin_2015}. The likelihood algorithm is validated over a theoretical grid by splitting the full set of simulations into two subsets: one used to estimate the covariance matrix and the other to evaluate the likelihood. We identified \ELiCAhybrid{}, which retains all cross-spectra and the WMAP-LFI auto-spectrum while marginalising over the HFI auto-spectra, as the most complete and unbiased likelihood configuration. Using the available cleaned dataset, from the low-$\ell$ E-mode power spectrum alone, we obtained $\tau = 0.0575_{-0.0058}^{+0.0048}$ at 68\% confidence level. 

When combined with the \Planck\ low-$\ell$ temperature likelihood \lowltt\, we found $\tau = 0.0587_{-0.0061}^{+0.0051}$ and $\ln(10^{10}A_{\mathrm{s}}) = 2.975_{-0.049}^{+0.048}$, consistent with previous analyses based on individual subsets of these data. Exploring the full parameter space of $\Lambda$CDM  with \CamSpec, and further including ACT DR6 + \Planck\ CMB lensing and DESI DR2 BAO measurements, we derived $\tau = 0.0632_{-0.0062}^{+0.0051}$ at 68\% confidence level. Adopting the same likelihood configuration, we consider the $\Lambda$CDM $+\sum m_\nu$ extension and place an upper bound on the total neutrino mass of $\sum m_\nu < 0.069,\mathrm{eV}$ at 95\% confidence level. Moreover, considering the $\Lambda$CDM +$N_{\mathrm{eff}}$ extension we find a constraint on the effective number of relativistic species of $N_{\mathrm{eff}} = 3.21 \pm 0.17$ at 68\% confidence level, in good agreement with the standard model prediction of $N_{\mathrm{eff}} = 3.044$.

Our results confirm that the value of $\tau$ favoured by all available large-scale satellite polarisation data remains around $0.06$. Assuming the standard late reionisation model, this excludes the higher values ($\tau \sim 0.09$) that would be required if the unexpectedly massive and evolved galaxies observed by the James Webb Space Telescope at $z > 10$ were efficient sources of ionising radiation, implying an earlier onset of reionisation. Similarly, independent analyses of DESI DR2 data have suggested scenarios with an increased optical depth in the context of the $\Lambda$CDM model. The constraints presented here provide a robust baseline against which such hypotheses can be tested.

Looking ahead, \elica~offers a framework that can be readily combined with complementary astrophysical probes of the Epoch of Reionisation — such as 21 cm observations, Lyman-$\alpha$ forest measurements, and high-redshift galaxy surveys — to constrain the reionisation history further. Moreover, the methodology presented here lays the groundwork for the analysis of future satellite missions such as LiteBIRD \citep{litebirdcollaboration2022ProbingCosmicInflationLiteBIRDa}, which will measure large-scale polarisation with unprecedented sensitivity.

\section*{Acknowledgements}

The authors thank L.P.L.~Colombo and M.~Gerbino for their valuable comments and discussions. Some of the results in this paper have been derived using the following packages: \texttt{camb} \citep{Lewis:1999bs, Howlett:2012mh}, \texttt{healpy} \citep{Gorski2005,zonca2009}, \texttt{Matplotlib} \citep{hunterMatplotlib2DGraphics2007}, \texttt{SciPy} \citep{virtanen2020SciPyFundamentalAlgorithmsscientific}, and \texttt{NumPy} \citep{harrisArrayProgrammingNumPy2020}. We acknowledge the financial support from the INFN InDark initiative and from the COSMOS network through the ASI (Italian Space Agency) Grants 2016-24-H.0 and 2016-24-H.1-2018. This work has also received funding by the European Union’s Horizon 2020 research and innovation  program under grant agreement no. 101007633 CMB-Inflate. P.C. and M.L. are funded by the European Union (ERC, RELiCS, project number 101116027). Views and opinions expressed are however those of the authors only and do not necessarily reflect those of the European Union or the European Research Council Executive Agency. Neither the European Union nor the granting authority can be held responsible for them. G.G. acknowledges support by the MUR PRIN2022 Project “BROWSEPOL: Beyond standaRd mOdel With coSmic microwavE background POLarization”-2022EJNZ53 financed by the European Union - Next Generation EU. We acknowledge CINECA for the availability of high performance computing resources and support through the CINECA-INFN agreement.

\bibliographystyle{aat}
\bibliography{bibliography,Planck_bib}

@ARTICLE{planck2016-l01,
author = {{\sorthelp{Planck Collaboration 2018A}}{Planck Collaboration I}},
title = "{\textit{Planck} 2018 results. I. Overview, and the cosmological
legacy of \textit{Planck}}",
journal = {\aap},
archivePrefix = "arXiv",
eprint = {1807.06205},
year = 2020,
volume = 641,
pages = {A1},
doi = {10.1051/0004-6361/201833880}
}

@ARTICLE{planck2016-l02,
author = {{\sorthelp{Planck Collaboration 2018B}}{Planck Collaboration II}},
title = "{\textit{Planck} 2018 results. II. Low Frequency Instrument data
 processing}",
journal = {\aap},
archivePrefix = "arXiv",
eprint = {1807.06206},
year = 2020,
volume = 641,
pages = {A2},
doi = {10.1051/0004-6361/201833293}
}

@ARTICLE{planck2016-l03, 
author = {{\sorthelp{Planck Collaboration 2018C}}{Planck Collaboration III}},
title = "{\textit{Planck} 2018 results. III. High Frequency Instrument data
 processing and frequency maps}",
journal = {\aap},
archivePrefix = "arXiv",
eprint = {1807.06207},
year = 2020,
volume = 641,
pages = {A3},
doi = {10.1051/0004-6361/201832909}
}

@ARTICLE{planck2016-l05, 
author = {{\sorthelp{Planck Collaboration 2018E}}{Planck Collaboration V}},
title = "{\textit{Planck} 2018 results. V. Power spectra and likelihoods}",
journal = {\aap},
archivePrefix = "arXiv",
eprint = {1907.12875},
year = 2020,
volume = 641,
pages = {A5},
doi = {10.1051/0004-6361/201936386}
}

@ARTICLE{planck2020-LVII,
author = {{\sorthelp{Planck Collaboration IntZZG}}{Planck Collaboration Int.
 LVII}},
title = "{\textit{Planck} intermediate results. LVII. NPIPE: Joint \Planck\ LFI
 and HFI data processing}",
journal = {\aap},
archivePrefix = "arXiv",
eprint = {2007.04997},
year = 2020,
volume = 643,
pages = {42},
doi = {10.1051/0004-6361/202038073},
}

@ARTICLE{zonca2009,
author = {{Zonca}, A. and {Franceschet}, C. and {Battaglia}, P. and {Villa}, F.
 and {Mennella}, A. and {D'Arcangelo}, O. and {Silvestri}, R. and
 {Bersanelli}, M. and {Artal}, E. and {Butler}, R.~C. and {Cuttaia}, F. and
 {Davis}, R.~J. and {Galeotta}, S. and {Hughes}, N. and {Jukkala}, P. and
 {Kilpi{\"a}}, {V.-H.} and {Laaninen}, M. and {Mandolesi}, N. and
 {Maris}, M. and {Mendes}, L. and {Sandri}, M. and {Terenzi}, L. and
 {Tuovinen}, J. and {Varis}, J. and {Wilkinson}, A.},
title = "{Planck-LFI radiometers' spectral response}",
journal = {Journal of Instrumentation},
archivePrefix = "arXiv",
eprint = {1001.4589},
primaryClass = "astro-ph.IM",
year = 2009,
month = dec,
volume = 4,
pages = {2010},
doi = {10.1088/1748-0221/4/12/T12010},
adsurl = {http://adsabs.harvard.edu/abs/2009JInst...4T2010Z}
}

@ARTICLE{keskitalo2010,
author = {{Keskitalo}, R. and {Ashdown}, M.~A.~J. and {Cabella}, P. and 
 {Kisner}, T. and {Poutanen}, T. and {Stompor}, R. and {Bartlett}, J.~G. and 
 {Borrill}, J. and {Cantalupo}, C. and {de Gasperis}, G. and 
 {de Rosa}, A. and {de Troia}, G. and {Eriksen}, H.~K. and {Finelli}, F. and 
 {G{\'o}rski}, K.~M. and {Gruppuso}, A. and {Hivon}, E. and {Jaffe}, A. and 
 {Keih{\"a}nen}, E. and {Kurki-Suonio}, H. and {Lawrence}, C.~R. and 
 {Natoli}, P. and {Paci}, F. and {Polenta}, G. and {Rocha}, G.},
title = "{Residual noise covariance for Planck low-resolution data analysis}",
journal = {\aap},
archivePrefix = "arXiv",
eprint = {0906.0175},
primaryClass = "astro-ph.CO",
year = 2010,
month = nov,
volume = 522,
eid = {A94},
pages = {A94},
doi = {10.1051/0004-6361/200912606},
adsurl = {http://adsabs.harvard.edu/abs/2010A%26A...522A..94K}
}

@ARTICLE{bennett2012,
author = {{Bennett}, C.~L. and {Larson}, D. and {Weiland}, J.~L. and {Jarosik},
 N. and {Hinshaw}, G. and {Odegard}, N. and {Smith}, K.~M. and {Hill}, R.~S.
 and {Gold}, B. and {Halpern}, M. and {Komatsu}, E. and {Nolta}, M.~R. and 
 {Page}, L. and {Spergel}, D.~N. and {Wollack}, E. and {Dunkley}, J. and 
 {Kogut}, A. and {Limon}, M. and {Meyer}, S.~S. and {Tucker}, G.~S. and 
 {Wright}, E.~L.},
title = "{Nine-year Wilkinson Microwave Anisotropy Probe (WMAP) Observations:
 Final Maps and Results}",
journal = {\apjs},
archivePrefix = "arXiv",
eprint = {1212.5225},
primaryClass = "astro-ph.CO",
year = 2013,
month = oct,
volume = 208,
eid = {20},
pages = {20},
doi = {10.1088/0067-0049/208/2/20},
adsurl = {http://adsabs.harvard.edu/abs/2013ApJS..208...20B}
}

@ARTICLE{gorski2005,
author = {{G{\'o}rski}, K.~M. and {Hivon}, E. and {Banday}, A.~J. and
 {Wandelt}, B.~D. and {Hansen}, F.~K. and {Reinecke}, M. and
 {Bartelmann}, M.},
title = "{HEALPix: A Framework for High-Resolution Discretization and Fast
 Analysis of Data Distributed on the Sphere}",
journal = {\apj},
archivePrefix = "arXiv",
eprint = {astro-ph/0409513},
year = 2005,
month = apr,
volume = 622,
pages = {759-771},
doi = {10.1086/427976},
adsurl = {http://adsabs.harvard.edu/abs/2005ApJ...622..759G}
}

@article{Hamimeche_2008,
   title={Likelihood analysis of CMB temperature and polarization power spectra},
   volume={77},
   ISSN={1550-2368},
   url={http://dx.doi.org/10.1103/PhysRevD.77.103013},
   DOI={10.1103/physrevd.77.103013},
   number={10},
   journal={Physical Review D},
   publisher={American Physical Society (APS)},
   author={Hamimeche, Samira and Lewis, Antony},
   year={2008},
   month=may }

@article{Mangilli_2015,
   title={Large-scale cosmic microwave background temperature and polarization cross-spectra likelihoods},
   volume={453},
   ISSN={1365-2966},
   url={http://dx.doi.org/10.1093/mnras/stv1733},
   DOI={10.1093/mnras/stv1733},
   number={3},
   journal={Monthly Notices of the Royal Astronomical Society},
   publisher={Oxford University Press (OUP)},
   author={Mangilli, A. and Plaszczynski, S. and Tristram, M.},
   year={2015},
   month=sep, pages={3175–3190} }

@phdthesis{vanneste:tel-02426412,
  TITLE = {{Constraints on primordial gravitational waves from the large scales CMB data}},
  AUTHOR = {Vanneste, Sylvain},
  URL = {https://theses.hal.science/tel-02426412},
  NUMBER = {2019SACLS314},
  SCHOOL = {{Universit{\'e} Paris Saclay (COmUE)}},
  YEAR = {2019},
  MONTH = Sep,
  TYPE = {Theses},
  HAL_ID = {tel-02426412},
  HAL_VERSION = {v1},
}

@article{Torrado_2021,
   title={Cobaya: code for Bayesian analysis of hierarchical physical models},
   volume={2021},
   ISSN={1475-7516},
   url={http://dx.doi.org/10.1088/1475-7516/2021/05/057},
   DOI={10.1088/1475-7516/2021/05/057},
   number={05},
   journal={Journal of Cosmology and Astroparticle Physics},
   publisher={IOP Publishing},
   author={Torrado, Jesús and Lewis, Antony},
   year={2021},
   month=may, pages={057} }

@article{Pagano:2019tci,
    author = "Pagano, L. and Delouis, J. -M. and Mottet, S. and Puget, J. -L. and Vibert, L.",
    title = "{Reionization optical depth determination from Planck HFI data with ten percent accuracy}",
    eprint = "1908.09856",
    archivePrefix = "arXiv",
    primaryClass = "astro-ph.CO",
    doi = "10.1051/0004-6361/201936630",
    journal = "Astron. Astrophys.",
    volume = "635",
    pages = "A99",
    year = "2020"
}

@article{Natale:2020owc,
    author = "Natale, U. and Pagano, L. and Lattanzi, M. and Migliaccio, M. and Colombo, L. P. and Gruppuso, A. and Natoli, P. and Polenta, G.",
    title = "{A novel CMB polarization likelihood package for large angular scales built from combined WMAP and Planck LFI legacy maps}",
    eprint = "2005.05600",
    archivePrefix = "arXiv",
    primaryClass = "astro-ph.CO",
    doi = "10.1051/0004-6361/202038508",
    journal = "Astron. Astrophys.",
    volume = "644",
    pages = "A32",
    year = "2020"
}

@article{Lattanzi:2016dzq,
    author = "Lattanzi, Massimiliano and Burigana, Carlo and Gerbino, Martina and Gruppuso, Alessandro and Mandolesi, Nazzareno and Natoli, Paolo and Polenta, Gianluca and Salvati, Laura and Trombetti, Tiziana",
    title = "{On the impact of large angle CMB polarization data on cosmological parameters}",
    eprint = "1611.01123",
    archivePrefix = "arXiv",
    primaryClass = "astro-ph.CO",
    doi = "10.1088/1475-7516/2017/02/041",
    journal = "JCAP",
    volume = "02",
    pages = "041",
    year = "2017"
}

@article{Paradiso:2022fky,
    author = "Paradiso, S. and others",
    title = "{BEYONDPLANCK - XII. Cosmological parameter constraints with end-to-end error propagation}",
    eprint = "2205.10104",
    archivePrefix = "arXiv",
    primaryClass = "astro-ph.CO",
    doi = "10.1051/0004-6361/202244060",
    journal = "Astron. Astrophys.",
    volume = "675",
    pages = "A12",
    year = "2023"
}

@article{Hinshaw2013,
    author = "Hinshaw, G. and others",
    collaboration = "WMAP",
    title = "{Nine-Year Wilkinson Microwave Anisotropy Probe (WMAP) Observations: Cosmological Parameter Results}",
    eprint = "1212.5226",
    archivePrefix = "arXiv",
    primaryClass = "astro-ph.CO",
    doi = "10.1088/0067-0049/208/2/19",
    journal = "Astrophys. J. Suppl.",
    volume = "208",
    pages = "19",
    year = "2013"
}

@article{galloni2025,
    author = "Galloni, Giacomo and Campeti, Paolo and Pagano, Luca and Gerbino, Martina and Lattanzi, Massimiliano and Natolia, Paolo",
    title = "{Accurate and efficient likelihood modeling for large-scale CMB data}",
    eprint = "2505.24829",
    archivePrefix = "arXiv",
    primaryClass = "astro-ph.CO",
    doi = "10.1088/1475-7516/2025/12/052",
    journal = "JCAP",
    volume = "12",
    pages = "052",
    year = "2025",
}

@article{Sellentin_2015,
   title={Parameter inference with estimated covariance matrices},
   volume={456},
   ISSN={1745-3933},
   url={http://dx.doi.org/10.1093/mnrasl/slv190},
   DOI={10.1093/mnrasl/slv190},
   number={1},
   journal={Monthly Notices of the Royal Astronomical Society: Letters},
   publisher={Oxford University Press (OUP)},
   author={Sellentin, Elena and Heavens, Alan F.},
   year={2015},
   month=dec, pages={L132–L136} }

@article{Lewis:1999bs,
      author         = "Lewis, Antony and Challinor, Anthony and Lasenby,
                        Anthony",
      title          = "{Efficient computation of CMB anisotropies in closed FRW
                        models}",
      journal        = "\apj",
      volume         = "538",
      year           = "2000",
      pages          = "473-476",
      doi            = "10.1086/309179",
      eprint         = "astro-ph/9911177",
      archivePrefix  = "arXiv",
      primaryClass   = "astro-ph",
      SLACcitation   = "%%CITATION = ASTRO-PH/9911177;%%"
}

@article{Howlett:2012mh,
      author         = "Howlett, Cullan and Lewis, Antony and Hall, Alex and
                        Challinor, Anthony",
      title          = "{CMB power spectrum parameter degeneracies in the era of
                        precision cosmology}",
      journal        = "\jcap",
      volume         = "1204",
      year           = "2012",
      pages          = "027",
      doi            = "10.1088/1475-7516/2012/04/027",
      eprint         = "1201.3654",
      archivePrefix  = "arXiv",
      primaryClass   = "astro-ph.CO",
      SLACcitation   = "%%CITATION = ARXIV:1201.3654;%%"
}

@article{DESI:2025zpo,
    author = "Abdul Karim, M. and others",
    collaboration = "DESI",
    title = "{DESI DR2 results. I. Baryon acoustic oscillations from the Lyman alpha forest}",
    eprint = "2503.14739",
    archivePrefix = "arXiv",
    primaryClass = "astro-ph.CO",
    reportNumber = "FERMILAB-PUB-25-0167-PPD",
    doi = "10.1103/2wwn-xjm5",
    journal = "Phys. Rev. D",
    volume = "112",
    number = "8",
    pages = "083514",
    year = "2025"
}

@article{DESI:2025zgx,
    author = "Abdul Karim, M. and others",
    collaboration = "DESI",
    title = "{DESI DR2 results. II. Measurements of baryon acoustic oscillations and cosmological constraints}",
    eprint = "2503.14738",
    archivePrefix = "arXiv",
    primaryClass = "astro-ph.CO",
    reportNumber = "FERMILAB-PUB-25-0169-PPD",
    doi = "10.1103/tr6y-kpc6",
    journal = "Phys. Rev. D",
    volume = "112",
    number = "8",
    pages = "083515",
    year = "2025"
}

@article{Rosenberg_2022,
   title={CMB power spectra and cosmological parameters from Planck PR4 with CamSpec},
   volume={517},
   ISSN={1365-2966},
   url={http://dx.doi.org/10.1093/mnras/stac2744},
   DOI={10.1093/mnras/stac2744},
   number={3},
   journal={Monthly Notices of the Royal Astronomical Society},
   publisher={Oxford University Press (OUP)},
   author={Rosenberg, Erik and Gratton, Steven and Efstathiou, George},
   year={2022},
   month=sep, pages={4620–4636} }

@article{Efstathiou_2021,
   title={A Detailed Description of the CAMSPEC Likelihood Pipeline and a Reanalysis of the Planck High Frequency Maps},
   volume={4},
   ISSN={2565-6120},
   url={http://dx.doi.org/10.21105/astro.1910.00483},
   DOI={10.21105/astro.1910.00483},
   number={1},
   journal={The Open Journal of Astrophysics},
   publisher={Maynooth University},
   author={Efstathiou, George and Gratton, Steven},
   year={2021},
   month=aug }

@article{Delouis:2019bub,
    author = "Delouis, J. -M. and Pagano, L. and Mottet, S. and Puget, J. -L. and Vibert, L.",
    title = "{SRoll2: an improved mapmaking approach to reduce large-scale systematic effects in the Planck High Frequency Instrument legacy maps}",
    eprint = "1901.11386",
    archivePrefix = "arXiv",
    primaryClass = "astro-ph.CO",
    doi = "10.1051/0004-6361/201834882",
    journal = "Astron. Astrophys.",
    volume = "629",
    pages = "A38",
    year = "2019"
}

@article{Tegmark:2001zv,
    author = "Tegmark, Max and de Oliveira-Costa, Angelica",
    title = "{How to measure CMB polarization power spectra without losing information}",
    eprint = "astro-ph/0012120",
    archivePrefix = "arXiv",
    doi = "10.1103/PhysRevD.64.063001",
    journal = "Phys. Rev. D",
    volume = "64",
    pages = "063001",
    year = "2001"
}

@article{Tegmark:1996qt,
    author = "Tegmark, Max",
    title = "{How to measure CMB power spectra without losing information}",
    eprint = "astro-ph/9611174",
    archivePrefix = "arXiv",
    reportNumber = "IASSNS-AST-96-63",
    doi = "10.1103/PhysRevD.55.5895",
    journal = "Phys. Rev. D",
    volume = "55",
    pages = "5895--5907",
    year = "1997"
}

@article{Gunn:1965hd,
      author         = "Gunn, James E. and Peterson, Bruce A.",
      title          = "{On the Density of Neutral Hydrogen in Intergalactic
                        Space}",
      journal        = "Astrophys. J.",
      volume         = "142",
      year           = "1965",
      pages          = "1633",
      doi            = "10.1086/148444",
      SLACcitation   = "%%CITATION = ASJOA,142,1633;%%"
}

@article{Montero-Camacho:2024dzs,
    author = "Montero-Camacho, Paulo and Li, Yin and Cranmer, Miles",
    title = "{Five parameters are all you need (in $\Lambda$CDM)}",
    eprint = "2405.13680",
    archivePrefix = "arXiv",
    primaryClass = "astro-ph.CO",
    month = "5",
    year = "2024",
    journal = "arXiv e-prints",
}

@article{Douspis:2015nca,
    author = "Douspis, Marian and Aghanim, Nabila and Ili{\'c}, St{\'e}phane and Langer, Mathieu",
    title = "{A new parameterization of the reionisation history}",
    eprint = "1509.02785",
    archivePrefix = "arXiv",
    primaryClass = "astro-ph.CO",
    doi = "10.1051/0004-6361/201526543",
    journal = "Astron. Astrophys.",
    volume = "580",
    pages = "L4",
    year = "2015"
}

@misc{kageura2026,
      title={A New Constraint on the Optical Depth from the Reionization History Independent of CMB Large-Scale E-Mode Polarization}, 
      author={Yuta Kageura and Masami Ouchi and Fumihiro Naokawa and Hiroya Umeda and Akinori Matsumoto and Yuichi Harikane and Minami Nakane and Tran Thi Thai},
      year={2026},
      eprint={2601.09644},
      archivePrefix={arXiv},
      primaryClass={astro-ph.CO},
      url={https://arxiv.org/abs/2601.09644}, 
}

@article{Tristram:2023haj,
    author = "Tristram, M. and others",
    title = "{Cosmological parameters derived from the final Planck data release (PR4)}",
    eprint = "2309.10034",
    archivePrefix = "arXiv",
    primaryClass = "astro-ph.CO",
    doi = "10.1051/0004-6361/202348015",
    journal = "Astron. Astrophys.",
    volume = "682",
    pages = "A37",
    year = "2024"
}

@article{Ilic:2025idl,
    author = "Ilic, S. and others",
    title = "{Reconstructing the epoch of reionisation with Planck PR4}",
    eprint = "2504.13254",
    archivePrefix = "arXiv",
    primaryClass = "astro-ph.CO",
    doi = "10.1051/0004-6361/202555196",
    journal = "Astron. Astrophys.",
    volume = "700",
    pages = "A26",
    year = "2025"
}

@article{Finkbeiner:1999aq,
    author = "Finkbeiner, Douglas P. and Davis, Marc and Schlegel, David J.",
    title = "{Extrapolation of galactic dust emission at 100 microns to CMBR frequencies using FIRAS}",
    eprint = "astro-ph/9905128",
    archivePrefix = "arXiv",
    doi = "10.1086/307852",
    journal = "Astrophys. J.",
    volume = "524",
    pages = "867--886",
    year = "1999"
}

@article{CLASS:2025khf,
    author = "Li, Yunyang and others",
    collaboration = "CLASS",
    title = "{A Measurement of the Largest-scale CMB E-mode Polarization with CLASS}",
    eprint = "2501.11904",
    archivePrefix = "arXiv",
    primaryClass = "astro-ph.CO",
    doi = "10.3847/1538-4357/adc723",
    journal = "Astrophys. J.",
    volume = "986",
    number = "2",
    pages = "111",
    year = "2025"
}

@article{Eimer:2023esh,
    author = "Eimer, Joseph R. and others",
    title = "{CLASS Angular Power Spectra and Map-component Analysis for 40 GHz Observations through 2022}",
    eprint = "2309.00675",
    archivePrefix = "arXiv",
    primaryClass = "astro-ph.CO",
    doi = "10.3847/1538-4357/ad1abf",
    journal = "Astrophys. J.",
    volume = "963",
    number = "2",
    pages = "92",
    year = "2024"
}

@article{Carron:2022eyg,
    author = "Carron, Julien and Mirmelstein, Mark and Lewis, Antony",
    title = "{CMB lensing from Planck PR4~maps}",
    eprint = "2206.07773",
    archivePrefix = "arXiv",
    primaryClass = "astro-ph.CO",
    doi = "10.1088/1475-7516/2022/09/039",
    journal = "JCAP",
    volume = "09",
    pages = "039",
    year = "2022"
}

@article{louis2025,
    title={The Atacama Cosmology Telescope: DR6 Power Spectra, Likelihoods and $\Lambda$CDM Parameters}, 
    author={Thibaut Louis and Adrien La Posta and Zachary Atkins and Hidde T. Jense and Irene Abril-Cabezas and Graeme E. Addison and Peter A. R. Ade and Simone Aiola and Tommy Alford and David Alonso and Mandana Amiri and Rui An and Jason E. Austermann and Eleonora Barbavara and Nicholas Battaglia and Elia Stefano Battistelli and James A. Beall and Rachel Bean and Ali Beheshti and Benjamin Beringue and Tanay Bhandarkar and Emily Biermann and Boris Bolliet and J Richard Bond and Erminia Calabrese and Valentina Capalbo and Felipe Carrero and Shi-Fan Chen and Grace Chesmore and Hsiao-mei Cho and Steve K. Choi and Susan E. Clark and Nicholas F. Cothard and Kevin Coughlin and William Coulton and Devin Crichton and Kevin T. Crowley and Omar Darwish and Mark J. Devlin and Simon Dicker and Cody J. Duell and Shannon M. Duff and Adriaan J. Duivenvoorden and Jo Dunkley and Rolando Dunner and Carmen Embil Villagra and Max Fankhanel and Gerrit S. Farren and Simone Ferraro and Allen Foster and Rodrigo Freundt and Brittany Fuzia and Patricio A. Gallardo and Xavier Garrido and Martina Gerbino and Serena Giardiello and Ajay Gill and Jahmour Givans and Vera Gluscevic and Samuel Goldstein and Joseph E. Golec and Yulin Gong and Yilun Guan and Mark Halpern and Ian Harrison and Matthew Hasselfield and Erin Healy and Shawn Henderson and Brandon Hensley and Carlos Hervías-Caimapo and J. Colin Hill and Gene C. Hilton and Matt Hilton and Adam D. Hincks and Renée Hložek and Shuay-Pwu Patty Ho and John Hood and Erika Hornecker and Zachary B. Huber and Johannes Hubmayr and Kevin M. Huffenberger and John P. Hughes and Margaret Ikape and Kent Irwin and Giovanni Isopi and Neha Joshi and Ben Keller and Joshua Kim and Kenda Knowles and Brian J. Koopman and Arthur Kosowsky and Darby Kramer and Aleksandra Kusiak and Alex Lague and Victoria Lakey and Eunseong Lee and Yaqiong Li and Zack Li and Michele Limon and Martine Lokken and Marius Lungu and Niall MacCrann and Amanda MacInnis and Mathew S. Madhavacheril and Diego Maldonado and Felipe Maldonado and Maya Mallaby-Kay and Gabriela A. Marques and Joshiwa van Marrewijk and Fiona McCarthy and Jeff McMahon and Yogesh Mehta and Felipe Menanteau and Kavilan Moodley and Thomas W. Morris and Tony Mroczkowski and Sigurd Naess and Toshiya Namikawa and Federico Nati and Simran K. Nerval and Laura Newburgh and Andrina Nicola and Michael D. Niemack and Michael R. Nolta and John Orlowski-Scherer and Luca Pagano and Lyman A. Page and Shivam Pandey and Bruce Partridge and Karen Perez Sarmiento and Heather Prince and Roberto Puddu and Frank J. Qu and Damien C. Ragavan and Bernardita Ried Guachalla and Keir K. Rogers and Felipe Rojas and Tai Sakuma and Emmanuel Schaan and Benjamin L. Schmitt and Neelima Sehgal and Shabbir Shaikh and Blake D. Sherwin and Carlos Sierra and Jon Sievers and Cristóbal Sifón and Sara Simon and Rita Sonka and David N. Spergel and Suzanne T. Staggs and Emilie Storer and Kristen Surrao and Eric R. Switzer and Niklas Tampier and Robert Thornton and Hy Trac and Carole Tucker and Joel Ullom and Leila R. Vale and Alexander Van Engelen and Jeff Van Lanen and Cristian Vargas and Eve M. Vavagiakis and Kasey Wagoner and Yuhan Wang and Lukas Wenzl and Edward J. Wollack and Kaiwen Zheng},
    collaboration = "Atacama Cosmology Telescope",
    eprint = "2503.14452",
    archivePrefix = "arXiv",
    primaryClass = "astro-ph.CO",
    doi = "10.1088/1475-7516/2025/11/062",
    journal = "JCAP",
    volume = "11",
    pages = "062",
    year = "2025"
}

@article{calabrese2025,
    title={The Atacama Cosmology Telescope: DR6 Constraints on Extended Cosmological Models}, 
    author={Erminia Calabrese and J. Colin Hill and Hidde T. Jense and Adrien La Posta and Irene Abril-Cabezas and Graeme E. Addison and Peter A. R. Ade and Simone Aiola and Tommy Alford and David Alonso and Mandana Amiri and Rui An and Zachary Atkins and Jason E. Austermann and Eleonora Barbavara and Nicola Barbieri and Nicholas Battaglia and Elia Stefano Battistelli and James A. Beall and Rachel Bean and Ali Beheshti and Benjamin Beringue and Tanay Bhandarkar and Emily Biermann and Boris Bolliet and J Richard Bond and Valentina Capalbo and Felipe Carrero and Shi-Fan Chen and Grace Chesmore and Hsiao-mei Cho and Steve K. Choi and Susan E. Clark and Nicholas F. Cothard and Kevin Coughlin and William Coulton and Devin Crichton and Kevin T. Crowley and Omar Darwish and Mark J. Devlin and Simon Dicker and Cody J. Duell and Shannon M. Duff and Adriaan J. Duivenvoorden and Jo Dunkley and Rolando Dunner and Carmen Embil Villagra and Max Fankhanel and Gerrit S. Farren and Simone Ferraro and Allen Foster and Rodrigo Freundt and Brittany Fuzia and Patricio A. Gallardo and Xavier Garrido and Martina Gerbino and Serena Giardiello and Ajay Gill and Jahmour Givans and Vera Gluscevic and Samuel Goldstein and Joseph E. Golec and Yulin Gong and Yilun Guan and Mark Halpern and Ian Harrison and Matthew Hasselfield and Adam He and Erin Healy and Shawn Henderson and Brandon Hensley and Carlos Hervías-Caimapo and Gene C. Hilton and Matt Hilton and Adam D. Hincks and Renée Hložek and Shuay-Pwu Patty Ho and John Hood and Erika Hornecker and Zachary B. Huber and Johannes Hubmayr and Kevin M. Huffenberger and John P. Hughes and Margaret Ikape and Kent Irwin and Giovanni Isopi and Neha Joshi and Ben Keller and Joshua Kim and Kenda Knowles and Brian J. Koopman and Arthur Kosowsky and Darby Kramer and Aleksandra Kusiak and Alex Lague and Victoria Lakey and Massimiliano Lattanzi and Eunseong Lee and Yaqiong Li and Zack Li and Michele Limon and Martine Lokken and Thibaut Louis and Marius Lungu and Niall MacCrann and Amanda MacInnis and Mathew S. Madhavacheril and Diego Maldonado and Felipe Maldonado and Maya Mallaby-Kay and Gabriela A. Marques and Joshiwa van Marrewijk and Fiona McCarthy and Jeff McMahon and Yogesh Mehta and Felipe Menanteau and Kavilan Moodley and Thomas W. Morris and Tony Mroczkowski and Sigurd Naess and Toshiya Namikawa and Federico Nati and Simran K. Nerval and Laura Newburgh and Andrina Nicola and Michael D. Niemack and Michael R. Nolta and John Orlowski-Scherer and Luca Pagano and Lyman A. Page and Shivam Pandey and Bruce Partridge and Karen Perez Sarmiento and Heather Prince and Roberto Puddu and Frank J. Qu and Damien C. Ragavan and Bernardita Ried Guachalla and Keir K. Rogers and Felipe Rojas and Tai Sakuma and Emmanuel Schaan and Benjamin L. Schmitt and Neelima Sehgal and Shabbir Shaikh and Blake D. Sherwin and Carlos Sierra and Jon Sievers and Cristóbal Sifón and Sara Simon and Rita Sonka and David N. Spergel and Suzanne T. Staggs and Emilie Storer and Kristen Surrao and Eric R. Switzer and Niklas Tampier and Leander Thiele and Robert Thornton and Hy Trac and Carole Tucker and Joel Ullom and Leila R. Vale and Alexander Van Engelen and Jeff Van Lanen and Cristian Vargas and Eve M. Vavagiakis and Kasey Wagoner and Yuhan Wang and Lukas Wenzl and Edward J. Wollack and Kaiwen Zheng},
    collaboration = "Atacama Cosmology Telescope",
    eprint = "2503.14454",
    archivePrefix = "arXiv",
    primaryClass = "astro-ph.CO",
    doi = "10.1088/1475-7516/2025/11/063",
    journal = "JCAP",
    volume = "11",
    pages = "063",
    year = "2025",
}

@article{Lodha2025,
  title = {Extended dark energy analysis using DESI DR2 BAO measurements},
  author = {Lodha, K. and Calderon, R. and Matthewson, W. L. and Shafieloo, A. and Ishak, M. and Pan, J. and Garcia-Quintero, C. and Huterer, D. and Valogiannis, G. and Ure\~na-L\'opez, L. A. and Kamble, N. V. and Parkinson, D. and Kim, A. G. and Zhao, G. B. and Cervantes-Cota, J. L. and Rohlf, J. and Lozano-Rodr\'{\i}guez, F. and Rom\'an-Herrera, J. O. and Abdul-Karim, M. and Aguilar, J. and Ahlen, S. and Alves, O. and Andrade, U. and Armengaud, E. and Aviles, A. and Behera, J. and BenZvi, S. and Bianchi, D. and Brodzeller, A. and Brooks, D. and Burtin, E. and Canning, R. and Rosell, A. Carnero and Casas, L. and Castander, F. J. and Charles, M. and Chaussidon, E. and Chaves-Montero, J. and Chebat, D. and Claybaugh, T. and Cole, S. and Cuceu, A. and Dawson, K. S. and de la Macorra, A. and de Mattia, A. and Deiosso, N. and Demina, R. and Dey, Arjun and Dey, Biprateep and Ding, Z. and Doel, P. and Eisenstein, D. J. and Elbers, W. and Ferraro, S. and Font-Ribera, A. and Forero-Romero, J. E. and Garrison, Lehman H. and Gazta\~naga, E. and Gil-Mar\'{\i}n, H. and Gontcho, S. Gontcho A. and Gonzalez-Morales, A. X. and Gutierrez, G. and Guy, J. and Hahn, C. and Herbold, M. and Herrera-Alcantar, H. K. and Honscheid, K. and Howlett, C. and Juneau, S. and Kehoe, R. and Kirkby, D. and Kisner, T. and Kremin, A. and Lahav, O. and Lamman, C. and Landriau, M. and Le Guillou, L. and Leauthaud, A. and Levi, M. E. and Li, Q. and Magneville, C. and Manera, M. and Martini, P. and Meisner, A. and Mena-Fern\'andez, J. and Miquel, R. and Moustakas, J. and Santos, D. Mu\~noz and Mu\~noz-Guti\'errez, A. and Myers, A. D. and Nadathur, S. and Niz, G. and Noriega, H. E. and Paillas, E. and Palanque-Delabrouille, N. and Percival, W. J. and Pieri, Matthew M. and Poppett, C. and Prada, F. and P\'erez-Fern\'andez, A. and P\'erez-R\`afols, I. and Ram\'{\i}rez-P\'erez, C. and Rashkovetskyi, M. and Ravoux, C. and Ross, A. J. and Rossi, G. and Ruhlmann-Kleider, V. and Samushia, L. and Sanchez, E. and Schlegel, D. and Schubnell, M. and Seo, H. and Sinigaglia, F. and Sprayberry, D. and Tan, T. and Tarl\'e, G. and Taylor, P. and Turner, W. and Vargas-Maga\~na, M. and Walther, M. and Weaver, B. A. and Wolfson, M. and Y\`eche, C. and Zarrouk, P. and Zhou, R. and Zou, H.},
  collaboration = {DESI Collaboration},
  journal = {Phys. Rev. D},
  volume = {112},
  issue = {8},
  pages = {083511},
  numpages = {27},
  year = {2025},
  month = {Oct},
  publisher = {American Physical Society},
  doi = {10.1103/w4c6-1r5j},
  url = {https://link.aps.org/doi/10.1103/w4c6-1r5j}
}

@article{Sailer_2026,
   title={Addressing Tensions in Cosmology by an Increase in the Optical Depth to Reionization},
   volume={136},
   ISSN={1079-7114},
   url={http://dx.doi.org/10.1103/6r54-8lv4},
   DOI={10.1103/6r54-8lv4},
   number={8},
   journal={Physical Review Letters},
   publisher={American Physical Society (APS)},
   author={Sailer, Noah and Farren, Gerrit S. and Ferraro, Simone and White, Martin},
   year={2026},
   month=feb }

@article{deBelsunce:2021mec,
    author = "de Belsunce, Roger and Gratton, Steven and Coulton, William and Efstathiou, George",
    title = "{Inference of the optical depth to reionization from low multipole temperature and polarization Planck data}",
    eprint = "2103.14378",
    archivePrefix = "arXiv",
    primaryClass = "astro-ph.CO",
    doi = "10.1093/mnras/stab2215",
    journal = "Mon. Not. Roy. Astron. Soc.",
    volume = "507",
    number = "1",
    pages = "1072--1091",
    year = "2021"
}

@article{hunterMatplotlib2DGraphics2007,
  ids = {matplotlib},
  title = {Matplotlib: {{A 2D}} Graphics Environment},
  author = {Hunter, J. D.},
  year = {2007},
  journal = {Computing in Science \& Engineering},
  volume = {9},
  number = {3},
  pages = {90--95},
  publisher = {{IEEE COMPUTER SOC}},
  doi = {10.1109/MCSE.2007.55}
}

@article{virtanen2020SciPyFundamentalAlgorithmsscientific,
  title = {{{SciPy}} 1.0: {{Fundamental}} Algorithms for Scientific Computing in {{Python}}},
  shorttitle = {{{SciPy}} 1.0},
  author = {Virtanen, Pauli and Gommers, Ralf and Oliphant, Travis E. and Haberland, Matt and Reddy, Tyler and Cournapeau, David and Burovski, Evgeni and Peterson, Pearu and Weckesser, Warren and Bright, Jonathan and {van der Walt}, St{\'e}fan J. and Brett, Matthew and Wilson, Joshua and Millman, K. Jarrod and Mayorov, Nikolay and Nelson, Andrew R. J. and Jones, Eric and Kern, Robert and Larson, Eric and Carey, C. J. and Polat, {\.I}lhan and Feng, Yu and Moore, Eric W. and VanderPlas, Jake and Laxalde, Denis and Perktold, Josef and Cimrman, Robert and Henriksen, Ian and Quintero, E. A. and Harris, Charles R. and Archibald, Anne M. and Ribeiro, Ant{\^o}nio H. and Pedregosa, Fabian and {van Mulbregt}, Paul},
  year = {2020},
  month = mar,
  journal = {Nature Methods},
  volume = {17},
  number = {3},
  eprint = {1907.10121},
  pages = {261--272},
  publisher = {{Nature Publishing Group}},
  issn = {1548-7105},
  doi = {10.1038/s41592-019-0686-2},
  visitdate = {2022-12-28},
  copyright = {2020 The Author(s)},
  langid = {english}
}

@article{harrisArrayProgrammingNumPy2020,
  ids = {numpy},
  title = {Array Programming with {{NumPy}}},
  author = {Harris, Charles R. and Millman, K. Jarrod and {van der Walt}, St{\'e}fan J and Gommers, Ralf and Virtanen, Pauli and Cournapeau, David and Wieser, Eric and Taylor, Julian and Berg, Sebastian and Smith, Nathaniel J. and Kern, Robert and Picus, Matti and Hoyer, Stephan and {van Kerkwijk}, Marten H. and Brett, Matthew and Haldane, Allan and {Fern{\'a}ndez del R{\'i}o}, Jaime and Wiebe, Mark and Peterson, Pearu and {G{\'e}rard-Marchant}, Pierre and Sheppard, Kevin and Reddy, Tyler and Weckesser, Warren and Abbasi, Hameer and Gohlke, Christoph and Oliphant, Travis E.},
  year = {2020},
  journal = {Nature},
  volume = {585},
  eprint = {2006.10256},
  pages = {357--362},
  doi = {10.1038/s41586-020-2649-2}
}

@article{elbers2025,
    title={Constraints on Neutrino Physics from DESI DR2 BAO and DR1 Full Shape}, 
    author={W. Elbers and A. Aviles and H. E. Noriega and D. Chebat and A. Menegas and C. S. Frenk and C. Garcia-Quintero and D. Gonzalez and M. Ishak and O. Lahav and K. Naidoo and G. Niz and C. Yèche and M. Abdul-Karim and S. Ahlen and O. Alves and U. Andrade and E. Armengaud and J. Behera and S. BenZvi and D. Bianchi and S. Brieden and A. Brodzeller and D. Brooks and E. Burtin and R. Calderon and R. Canning and A. Carnero Rosell and L. Casas and F. J. Castander and M. Charles and E. Chaussidon and J. Chaves-Montero and T. Claybaugh and S. Cole and A. P. Cooper and A. Cuceu and K. S. Dawson and A. de la Macorra and A. de Mattia and N. Deiosso and A. Dey and B. Dey and Z. Ding and P. Doel and D. J. Eisenstein and S. Ferraro and A. Font-Ribera and J. E. Forero-Romero and L. H. Garrison and E. Gaztañaga and H. Gil-Marín and S. Gontcho A Gontcho and A. X. Gonzalez-Morales and G. Gutierrez and S. He and M. Herbold and H. K. Herrera-Alcantar and C. Howlett and D. Huterer and S. Juneau and R. Kehoe and D. Kirkby and T. Kisner and A. Kremin and C. Lamman and M. Landriau and L. Le Guillou and A. Leauthaud and M. E. Levi and Q. Li and K. Lodha and C. Magneville and M. Manera and P. Martini and W. L. Matthewson and A. Meisner and J. Mena-Fernández and R. Miquel and J. Moustakas and S. Nadathur and J. A. Newman and E. Paillas and N. Palanque-Delabrouille and W. J. Percival and M. M. Pieri and C. Poppett and F. Prada and I. Pérez-Ràfols and D. Rabinowitz and C. Ramírez-Pérez and M. Rashkovetskyi and C. Ravoux and H. Rivera-Morales and J. Rohlf and A. J. Ross and G. Rossi and V. Ruhlmann-Kleider and L. Samushia and E. Sanchez and D. Schlegel and M. Schubnell and H. Seo and F. Sinigaglia and D. Sprayberry and T. Tan and G. Tarlé and P. Taylor and W. Turner and M. Vargas-Magaña and L. Verde and M. Walther and B. A. Weaver and A. Whitford and M. Wolfson and P. Zarrouk and C. Zhao and R. Zhou and H. Zou},
    eprint = "2503.14744",
    archivePrefix = "arXiv",
    primaryClass = "astro-ph.CO",
    doi = "10.1103/w9pk-xsk7",
    journal = "Phys. Rev. D",
    volume = "112",
    number = "8",
    pages = "083513",
    year = "2025" 
}

@article{litebirdcollaboration2022ProbingCosmicInflationLiteBIRDa,
  title = {Probing {{Cosmic Inflation}} with the {{LiteBIRD Cosmic Microwave Background Polarization Survey}}},
  author = {Allys, E. and Arnold, K. and Aumont, J. and Aurlien, R. and Azzoni, S. and Baccigalupi, C. and Banday, A. J. and Banerji, R. and Barreiro, R. B. and Bartolo, N. and Bautista, L. and Beck, D. and Beckman, S. and Bersanelli, M. and Boulanger, F. and Brilenkov, M. and Bucher, M. and Calabrese, E. and Campeti, P. and Carones, A. and Casas, F. J. and Catalano, A. and Chan, V. and Cheung, K. and Chinone, Y. and Clark, S. E. and Columbro, F. and D'Alessandro, G. and {de Bernardis}, P. and {de Haan}, T. and {de la Hoz}, E. and De Petris, M. and Della Torre, S. and {Diego-Palazuelos}, P. and Dobbs, M. and Dotani, T. and Duval, J. M. and Elleflot, T. and Eriksen, H. K. and Errard, J. and {Essinger-Hileman}, T. and Finelli, F. and Flauger, R. and Franceschet, C. and Fuskeland, U. and Galloway, M. and Ganga, K. and Gerbino, M. and Gervasi, M. and {G{\'e}nova-Santos}, R. T. and Ghigna, T. and Giardiello, S. and Gjerl{\o}w, E. and Grain, J. and Grupp, F. and Gruppuso, A. and Gudmundsson, J. E. and Halverson, N. W. and Hargrave, P. and Hasebe, T. and Hasegawa, M. and Hazumi, M. and {Henrot-Versill{\'e}}, S. and Hensley, B. and Hergt, L. T. and Herman, D. and Hivon, E. and Hlozek, R. A. and Hornsby, A. L. and Hoshino, Y. and Hubmayr, J. and Ichiki, K. and Iida, T. and Imada, H. and Ishino, H. and Jaehnig, G. and Katayama, N. and Kato, A. and Keskitalo, R. and Kisner, T. and Kobayashi, Y. and Kogut, A. and Kohri, K. and Komatsu, E. and Komatsu, K. and Konishi, K. and Krachmalnicoff, N. and Kuo, C. L. and Lamagna, L. and Lattanzi, M. and Lee, A. T. and Leloup, C. and Levrier, F. and Linder, E. and Luzzi, G. and {Macias-Perez}, J. and Maciaszek, T. and Maffei, B. and Maino, D. and Mandelli, S. and {Mart{\'i}nez-Gonz{\'a}lez}, E. and Masi, S. and Massa, M. and Matarrese, S. and Matsuda, F. T. and Matsumura, T. and Mele, L. and Migliaccio, M. and Minami, Y. and Moggi, A. and Montgomery, J. and Montier, L. and Morgante, G. and Mot, B. and Nagano, Y. and Nagasaki, T. and Nagata, R. and Nakano, R. and Namikawa, T. and Nati, F. and Natoli, P. and Nerval, S. and Noviello, F. and Odagiri, K. and Oguri, S. and Ohsaki, H. and Pagano, L. and Paiella, A. and Paoletti, D. and Passerini, A. and Patanchon, G. and Piacentini, F. and Piat, M. and Polenta, G. and Poletti, D. and Prouv{\'e}, T. and Puglisi, G. and Rambaud, D. and Raum, C. and Realini, S. and Reinecke, M. and Remazeilles, M. and Ritacco, A. and Roudil, G. and {Rubino-Martin}, J. A. and Russell, M. and Sakurai, H. and Sakurai, Y. and Sasaki, M. and Scott, D. and Sekimoto, Y. and Shinozaki, K. and Shiraishi, M. and Shirron, P. and Signorelli, G. and Spinella, F. and Stever, S. and Stompor, R. and Sugiyama, S. and Sullivan, R. M. and Suzuki, A. and Svalheim, T. L. and Switzer, E. and Takaku, R. and Takakura, H. and Takase, Y. and Tartari, A. and Terao, Y. and Thermeau, J. and Thommesen, H. and Thompson, K. L. and Tomasi, M. and Tominaga, M. and Tristram, M. and Tsuji, M. and Tsujimoto, M. and Vacher, L. and Vielva, P. and Vittorio, N. and Wang, W. and Watanuki, K. and Wehus, I. K. and Weller, J. and Westbrook, B. and Wilms, J. and Wollack, E. J. and Yumoto, J. and Zannoni, M.},
  year = {2022},
  month = nov,
  journal = {Progress of Theoretical and Experimental Physics},
  eprint = {2202.02773},
  primaryclass = {astro-ph},
  pages = {ptac150},
  issn = {2050-3911},
  doi = {10.1093/ptep/ptac150},
  visitdate = {2022-12-21},
  archiveprefix = {arxiv}
}

@article{ACT:2023dou,
    author = "Qu, Frank J. and others",
    collaboration = "ACT",
    title = "{The Atacama Cosmology Telescope: A Measurement of the DR6 CMB Lensing Power Spectrum and Its Implications for Structure Growth}",
    eprint = "2304.05202",
    archivePrefix = "arXiv",
    primaryClass = "astro-ph.CO",
    reportNumber = "FERMILAB-PUB-23-237-PPD, FERMILAB-PUB-23-237-PPD",
    doi = "10.3847/1538-4357/acfe06",
    journal = "Astrophys. J.",
    volume = "962",
    number = "2",
    pages = "112",
    year = "2024"
}

@article{ACT:2023kun,
    author = "Madhavacheril, Mathew S. and others",
    collaboration = "ACT",
    title = "{The Atacama Cosmology Telescope: DR6 Gravitational Lensing Map and Cosmological Parameters}",
    eprint = "2304.05203",
    archivePrefix = "arXiv",
    primaryClass = "astro-ph.CO",
    reportNumber = "FERMILAB-PUB-23-206-PPD",
    doi = "10.3847/1538-4357/acff5f",
    journal = "Astrophys. J.",
    volume = "962",
    number = "2",
    pages = "113",
    year = "2024"
}

@article{munoz2024,
  author  = {Mu{\~n}oz, J. B. and Mirocha, J. and Chisholm, J. and Furlanetto, S. R. and Mason, C.},
  title   = {Reionization after JWST: a photon budget crisis?},
  journal = {Monthly Notices of the Royal Astronomical Society: Letters},
  volume  = {535},
  number  = {1},
  pages   = {L37--L43},
  year    = {2024},
  doi     = {10.1093/mnrasl/slae086}
}

@misc{tan2025,
      title={The Impact of Population III.1 Flash Reionization for CMB Polarization and Thomson Scattering Optical Depth}, 
      author={Jonathan C. Tan and Eiichiro Komatsu},
      year={2025},
      eprint={2510.19647},
      archivePrefix={arXiv},
      primaryClass={astro-ph.CO},
      url={https://arxiv.org/abs/2510.19647}, 
}

\appendix
\section{Review of QML algorithm} \label{sec:QML}
In this appendix, we summarise the main steps of the QML algorithm, used to estimate the CMB power spectra from pixel-space maps. The QML is unbiased and minimises variance by optimally combining signal and noise information \citep{Tegmark:1996qt}. Starting from a map $\mathbf{m}$, which contains contributions from both signal and noise, the pixel covariance matrix $\mathbf{C}$ is defined as:  
\begin{equation}
\mathbf{C} \equiv \langle \mathbf{m}\mathbf{m}^\mathrm{T} \rangle = \mathbf{N} + \mathbf{S},
\label{eq:covariance}
\end{equation}
where $\langle \cdot \rangle$ represents the ensemble average, $\mathbf{N}$ is the noise covariance and the signal covariance $\mathbf{S}$ is given as:  
\begin{equation}
\mathbf{S} \equiv \sum_\ell \, \mathbf{P}_{ij}^{\ell} \, C_\ell, 
\label{eq:signal_covariance}
\end{equation}
where $\mathbf{P}^{\ell}$ is the matrix of Legendre polynomials at a given multipole and $C_\ell$ encodes all the possible power spectra $TT$, $EE$, $BB$, $TE$, $TB$, and $EB$. 

For two generic polarisation maps $\mathbf{m}^A=[\mathbf{Q}^A,\mathbf{U}^A]$ and $\mathbf{m}^B=[\mathbf{Q}^B,\mathbf{U}^B]$ corresponding to channels $A$ and $B$, the QML estimate of the cross power spectrum is

\begin{equation}
\hat{C}^{AB}_{\ell} \equiv \mathbf{m}^{A\,\mathrm{T}}
 \mathbf{E}^\ell \, \mathbf{m}^B - b^{AB}_\ell \, ,
\end{equation}
where the symmetric matrix $\mathbf{E}^{\ell}$ is derived from the Fisher matrix $F_{\ell\ell}$ and the total pixel covariance  $\mathbf{C}$:
\begin{equation}
\mathbf{E}^{\ell} = \frac{1}{2} \, \, \frac{ \mathbf{C}^{-1}\mathbf{P}^{\ell} \mathbf{C}^{-1}  }{F_{\ell\ell}} .
\end{equation}

For cross-spectra, the noise bias $b^{AB}_{\ell}$ vanishes under the assumption of uncorrelated noise between channels. For the respective auto-spectra, the bias:

\begin{equation}
\label{eq:bl_bias}
b^{AA}_\ell = \mathrm{Tr}[ \mathbf{N}^A \mathbf{E}^\ell], \qquad
b^{BB}_\ell = \mathrm{Tr}[ \mathbf{N}^B \mathbf{E}^\ell],
\end{equation}
must be subtracted to obtain an unbiased estimate:
\begin{equation}
\hat{C}^{AA}_{\ell} \equiv \mathbf{m}^{A\,\mathrm{T}} \mathbf{E}^\ell \mathbf{m}^A ,
\qquad
\hat{C}^{BB}_{\ell} \equiv \mathbf{m}^{B\,\mathrm{T}} \mathbf{E}^\ell \mathbf{m}^B,
\end{equation}

where, $\mathbf{N}^A$ and $\mathbf{N}^B$ are the noise covariance matrices for channels $A$ and $B$, respectively. 

We note that the noise bias $\mathrm{Tr}[\mathbf{N} \mathbf{E}^\ell]$ (Eq.~\ref{eq:bl_bias}) differs between noise-only maps $\mathbf{m}=\mathbf{N}$ and those containing both CMB and noise $\mathbf{m}=\mathbf{S}+\mathbf{N}$. This is due to the algebra of the QML estimator as the matrix $\mathbf{E}^\ell$ is constructed from the covariance $\mathbf{C} = \mathbf{N}$ in the first case, and $\mathbf{C} = \mathbf{S} + \mathbf{N}$ in the second case. For this reason, after computing unbiased noise-only power spectra with the QML estimator, we re-bias them by adding the $b_\ell$ term estimated from the CMB+noise maps, rather than from noise-only realisations. 

\section{Alternative high-\texorpdfstring{$\ell$}{TEXT} likelihoods}\label{sec:altre_highl}
In this section, we summarise the main results obtained by testing \ELiCAhybrid{} with alternative high-$\ell$ likelihoods, keeping the 2018 low-$\ell$ temperature likelihood. In particular, we consider the \Plik\ likelihood\footnote{We used the \texttt{lite\_native} version.}, previously used in \citep{Natale:2020owc, Pagano:2019tci}. In those works, the use of the \Plik\ high-$\ell$ likelihood led to an increase in the value of $A_{\mathrm{s}}$ and pulled $\tau$ toward higher values. This trend is also observed in \ELiCAhybrid{}, as Tab.~\ref{tab:other_high_likelihoods}. In contrast, the \CamSpec\ likelihood favours a lower value of $A_{\mathrm{s}}$ \citep{Rosenberg_2022}, mitigating the dragging of $\tau$ when including high-$\ell$ data, as shown in Eq.~\ref{eq:ELiCA_CamSpec_constraints}. Moreover, we include the ACT high-$\ell$ likelihood, which combines the \Planck\ data on large-to-intermediate scales — at $\ell$ < 1000 in TT and $\ell$ < 600 in TE/EE and ACT measurements up to $\ell=8500$ \cite{louis2025}. Since this likelihood provides constraining power in a different region of the power spectrum compared to \Planck\ (particularly in polarisation), performing a joint fit to both datasets is of considerable interest.

\begin{table}[htbp]\label{tab:other_high_likelihoods}
\begingroup
\centering  
\caption{Constraints on $\tau$ and $A_{\mathrm{s}}$ obtained using different high-$\ell$ likelihoods. We report the 68\% confidence level.}
\label{tab:PTE_parameters}
\nointerlineskip
\vskip -3mm
\setbox\tablebox=\vbox{
   \newdimen\digitwidth
   \setbox0=\hbox{\rm 0}
   \digitwidth=\wd0
   \catcode`*=\active
   \def*{\kern\digitwidth}
   \newdimen\signwidth
   \setbox0=\hbox{+}
   \signwidth=\wd0
   \catcode`!=\active
   \def!{\kern\signwidth}
\halign{\hbox to 1.0in{#\leaderfil}\tabskip=1em&
  \hfil#\hfil\tabskip=1em&
  \hfil#\hfil\tabskip=0pt\cr
\noalign{\doubleline}
\noalign{\vskip - 2pt}
\omit\hfil Likelihood \hfil & \hfil $\tau$ \hfil & $\ln(10^{10} A_{\mathrm{s}})$ \cr
\noalign{\vskip 3pt\hrule\vskip 5pt}
\CamSpec      & $0.0581_{-0.0059}^{+0.0048}$    & $3.048_{-0.012}^{+0.011}$ \cr
\Plik         & $0.0592^{+0.0050}_{-0.0059}$    & $3.055^{+0.011}_{-0.012}$ \cr
\Hillipop     & $0.0581^{+0.0049}_{-0.0056}$    & $3.041 \pm 0.012$ \cr
\ACTplanckcut& $0.0609^{+0.0050}_{-0.0060}$    & $3.058^{+0.011}_{-0.012}$ \cr
\noalign{\vskip 6pt\hrule\vskip 4pt}}}
\centerline{\box\tablebox}
\endgroup
\end{table}

\section{Consistency with ACT DR6 results} \label{sec:ACT_DR6}
In this section, we summarise the main results obtained by testing \ELiCAhybrid{} using the analysis pipeline adopted in the ACT DR6 analysis \citep{louis2025, calabrese2025}, while replacing the DESI DR1 BAO measurements with the more recent DESI DR2 release. Therefore, the main difference with respect to the baseline analysis presented in Section \ref{sec:cosmology} is the use of the \ACTplanckcut{} likelihood in the high-$\ell$ region, which combines ACT measurements up to $\ell = 8500$ with \Planck\ data on large-to-intermediate angular scales. We further include the joint ACT DR6+ \Planck\ CMB lensing likelihood (\lensing{}, \citet{ACT:2023dou,ACT:2023kun,Carron:2022eyg}) and DESI DR2 BAO measurements (\BAO{}, \citet{DESI:2025zpo,DESI:2025zgx}). The resulting constraints are reported in Tables~\ref{tab:TAU_parameters_ACT}. The constraint on the amplitude of scalar perturbations is unchanged after including \lensing{} and \BAO{} likelihoods, while the optical depth constraint shifts. This is consistent with \citep{louis2025} (Fig.~44, Appendix H).

\begin{table}[htbp]
\begingroup
\centering 
\caption{Constraints on $\tau$ and $A_{\mathrm{s}}$ obtained using the \ACTplanckcut{} high-$\ell$ likelihood. We report the 68\% confidence level.}
\label{tab:TAU_parameters_ACT}
\nointerlineskip
\vskip -3mm
\setbox\tablebox=\vbox{
   \newdimen\digitwidth
   \setbox0=\hbox{\rm 0}
   \digitwidth=\wd0
   \catcode`*=\active
   \def*{\kern\digitwidth}
   \newdimen\signwidth
   \setbox0=\hbox{+}
   \signwidth=\wd0
   \catcode`!=\active
   \def!{\kern\signwidth}
\halign{\hbox to 1.5in{#\leaderfil}\tabskip=1em&
  \hfil#\hfil\tabskip=1em&
  \hfil#\hfil\tabskip=0pt\cr
\noalign{\doubleline}
\noalign{\vskip -1.5pt}
\omit\hfil Likelihood \hfil & \hfil $\tau$ \hfil & \hfil $\ln(10^{10} A_{\mathrm{s}}) $ \hfil \cr

\ACTplanckcut & $0.0609^{+0.0050}_{-0.0060}$ & $3.058^{+0.011}_{-0.012}$ \cr
\ACTplanckcut +\vtop{\hbox{\lensing}\hbox{+ \BAO}} & $0.0640^{+0.0053}_{-0.0060}$ & $3.057\pm 0.012$ \cr
\noalign{\vskip 6pt\hrule\vskip 4pt}}}
\centerline{\box\tablebox}
\endgroup
\end{table}
Using the \ACTplanckcut{} as high-$\ell$ likelihood, combined with lensing and DESI DR2 likelihood, we explore the $\Lambda$CDM+$\sum m_\nu$ and $\Lambda$CDM+$N_{\rm eff}$ extensions. As in Section \ref{sec:cosmology}, we assume three massive neutrino species.

\begin{table}[htbp]
\begingroup
\centering
\caption{Constraints on $\sum m_\nu$ and $N_{\rm eff}$ obtained using the \ACTplanckcut{} high-$\ell$ likelihood. We report the 95\% confidence level for the neutrino mass and 68 \% confidence level for the $N_{\rm eff}$.}
\label{tab:neutrino_parameters_ACT}
\nointerlineskip
\vskip -3mm
\setbox\tablebox=\vbox{
   \newdimen\digitwidth
   \setbox0=\hbox{\rm 0}
   \digitwidth=\wd0
   \catcode`*=\active
   \def*{\kern\digitwidth}
   \newdimen\signwidth
   \setbox0=\hbox{+}
   \signwidth=\wd0
   \catcode`!=\active
   \def!{\kern\signwidth}
\halign{\hbox to 1.5in{#\leaderfil}\tabskip=1em&
  \hfil#\hfil\tabskip=1em&
  \hfil#\hfil\tabskip=0pt\cr
\noalign{\doubleline}
\noalign{\vskip -0.5pt}
\omit\hfil Likelihood \hfil & \hfil $\sum m_\nu$ [eV] \hfil & \hfil $\mathrm{N_{eff}}$ \hfil \cr
\noalign{\vskip 3pt\hrule\vskip 5pt}

$\Lambda\mathrm{CDM} + \sum m_\nu$ & $<0.077$ & --- \cr
\noalign{\vskip 4pt}
$\Lambda\mathrm{CDM} + N_{\rm eff}$ & --- & $2.92\pm 0.12$ \cr
\noalign{\vskip 6pt\hrule\vskip 4pt}}}
\centerline{\box\tablebox}
\endgroup
\end{table}

\end{document}